\documentclass{article}

\usepackage{arxiv}

\usepackage[T1]{fontenc}    
\usepackage{hyperref}       
\usepackage{url}            
\usepackage{booktabs}       
\usepackage{amsfonts}       
\usepackage{nicefrac}       
\usepackage{microtype}      
\usepackage{graphicx}
\usepackage{natbib}
\usepackage{doi}

\usepackage{makecell}
\usepackage[utf8x]{inputenc}
\usepackage{tfrupee}
\usepackage{ragged2e}
\usepackage{array}
\usepackage{algorithm}        
\usepackage{algpseudocode}    
\usepackage{amsmath}          
\usepackage{microtype}        
\usepackage{etoolbox}         
\usepackage{longtable}
\usepackage{enumitem}
\usepackage{framed}

\title{Developing \textit{JEEVHITAA}: Designing and Evaluating a Human-Centered AI Ecosystem for Collective Care}
\renewcommand{\shorttitle}{Developing \textit{JEEVHITAA}}

\author{ \href{https://orcid.org/0000-0001-5787-4858}{\includegraphics[scale=0.06]{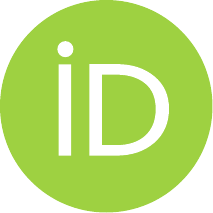}\hspace{1mm}Shyama Sastha~Krishnamoorthy Srinivasan}\\
	IIIT-Delhi\\
	New Delhi, India\\
	\texttt{shyamas@iiitd.ac.in} \\
    \And
    \href{https://orcid.org/0009-0006-6650-924X}
    {\includegraphics[scale=0.06]{orcid.pdf}\hspace{1mm}Harsh~Pala}\\
	IIIT-Delhi\\
	New Delhi, India\\
	\texttt{harsh24119@iiitd.ac.in} \\
    \And
    \href{https://orcid.org/0009-0000-4433-3454}
    {\includegraphics[scale=0.06]{orcid.pdf}\hspace{1mm}Shahnawaz}\\
	IIIT-Delhi\\
	New Delhi, India\\
	\texttt{shahnawaz25042@iiitd.ac.in} \\
    \And
    {\includegraphics[scale=0.06]{orcid.pdf}\hspace{1mm}Mohan~Kumar}\thanks{Both authors are advisors of this work.}\\
	RIT, NY\\
	Rochester, New York, US\\
	\texttt{mjkvcs@rit.edu} \\
	\And
	\href{https://orcid.org/0000-0003-2152-1027}{\includegraphics[scale=0.06]{orcid.pdf}\hspace{1mm}Pushpendra~Singh}\footnotemark[1]\\
	IIIT-Delhi\\
	New Delhi, India\\
	\texttt{psingh@iiitd.ac.in} \\
}

\hypersetup{
pdftitle={Developing JEEVHITAA: Designing and Evaluating a Human-Centered AI Ecosystem for Collective Care},
pdfsubject={cs.HC, cs.CY},
pdfauthor={Shyama Sastha~Krishnamoorthy Srinivasan, Harsh~Pala, Shahnawaz, Mohan~Kumar, Pushpendra~Singh},
pdfkeywords={Ecology of tracking, Personal Health Management, Collective Care},
}

\begin{document}
\maketitle

\begin{abstract}
	Health management is often a collective activity, involving family members and close relations in monitoring, interpreting, and coordinating care. Yet, most digital health technologies remain individual-centric, offering limited support for the distributed responsibilities and negotiated information sharing that characterize collective care. We present \textit{JEEVHITAA}, a human-centered AI ecosystem designed to address this gap. Drawing on existing literature on collective health practices, we identify key requirements and translate them into a canonical ontology and a versioned data schema that underpin a mobile ecosystem integrating wearable data, role-aware sharing, granular access controls, auditable histories, and AI-assisted health summarization and verification. We first evaluated core system primitives using 1,000 ontology-driven synthetic care circles to test data visibility and information-processing workflows. We then conducted a controlled lab study with 13 subject\textendash care-actor dyads (26 participants), combining task-based interactions, semi-structured interviews, the System Usability Scale, and component-level usability assessments. Findings indicate positive perceptions of usability and highlight the value of granular sharing and transparent information provenance. They also expose tensions between reciprocal transparency and surveillance, and between AI-mediated assistance and potentially intrusive notifications. We discuss implications for designing accountable, context-sensitive technologies that support collective care.
\end{abstract}

\keywords{Health Informatics \and Information Systems \and Collective Care}

\section{Introduction}

Health management is rarely an exclusively individual activity. Decisions about everyday health and wellbeing frequently involve family members, partners, and close companions who collectively monitor health indicators, interpret changes, exchange information, and coordinate appropriate responses. Research on family informatics has demonstrated how health tracking extends beyond individual self-monitoring, with family members sharing tracking responsibilities and interpreting health information in relation to one another's needs~\cite{10.1145/2998181.2998362}. Similarly, research on interpersonal informatics highlights how personal health management is situated within broader social relationships, where the availability, involvement, and influence of others shape how information is understood and acted upon~\cite{10.1145/3274396}. These practices are particularly relevant in the Indian context, where personal health technologies are frequently appropriated to support shared and collective care practices rather than solely individual self-management~\cite{srinivasan2026unpackingpersonalhealthinformatics}. Despite this reality, most consumer health applications and wearable ecosystems continue to organize their interfaces, data representations, and feedback mechanisms around an individual account holder. Consequently, the interpretation and sharing of health information across relationships often rely on informal workarounds, such as exchanging screenshots, sharing device credentials, or repeatedly relaying health updates via messaging platforms.

Extending personal health technologies to support collective care, however, involves more than providing another mechanism for sharing health data. Health information is relationally sensitive: the relevance of a particular measurement, the legitimacy of accessing it, and the responsibilities associated with interpreting it can differ across relationships and circumstances. Studies of health datafication have shown that tracking and sharing information can redistribute caregiving responsibilities while introducing emotional, interpersonal, and practical burdens~\cite{10.1145/2998181.2998303,10.1145/3715336.3735746}. Moreover, consent and disclosure preferences are not necessarily permanent decisions. Individuals may wish to expose different information to different people, or change access as their circumstances and relationships evolve~\cite{10.1145/2441776.2441838}. Research on dynamic privacy controls and intimate health-data sharing further demonstrates the importance of making information flows visible and allowing users to exercise meaningful control over sensitive data~\cite{10.1145/3613904.3642815,10.1145/3706598.3714089}. These findings suggest that collective health technologies require interaction mechanisms that account for relationship-specific responsibilities, differentiated information visibility, and the changing conditions under which health information is shared. Without such mechanisms, expanding access to health data may inadvertently create new expectations of accountability, unwanted monitoring, and additional coordination work.

The integration of artificial intelligence (AI) introduces another dimension to this challenge. Increasingly, health applications use intelligent systems to summarize personal measurements, generate recommendations, and help individuals interpret health-related information. In collective settings, these outputs may be consumed by multiple actors with different levels of health literacy, different responsibilities, and potentially different interpretations of the same information. An AI-generated recommendation, therefore, cannot be understood solely in terms of its usefulness to an individual recipient; its source, supporting evidence, contextual relevance, and implications for other members of the care circle also matter. Although explainability and interpretability are frequently presented as mechanisms for improving trust in AI, evidence suggests that making a model more interpretable does not necessarily produce appropriately calibrated trust or improve human decision-making~\cite{10.1145/3613904.3642780}. Human-AI interaction guidelines consequently emphasize the importance of communicating system capabilities, uncertainty, and supporting information in ways that enable users to evaluate rather than merely accept automated outputs~\cite{10.1145/3290605.3300233}. In health contexts, this motivates interfaces that expose data provenance, provide inspectable supporting references, and distinguish automated confidence assessments from independently verifiable evidence.

To investigate how these requirements can be translated into an interactive system, we developed \textit{JEEVHITAA} (Joint Engagement Ecosystem for Vital Health Insights, Tracking, and Actionable Assistance), a mobile human-centered AI ecosystem designed to support health information sharing and coordinated interpretation within small care circles. Rather than treating the individual account holder as the only relevant user, \textit{JEEVHITAA} accommodates two complementary perspectives: the \textit{Subject}, whose health information is collected and shared, and the \textit{Care Actor}, who receives selected information to provide support. The system brings together four principal interaction capabilities: (1) wearable-based health dashboards and contextualized summaries; (2) subject-controlled, metric-level sharing and role-specific visibility; (3) provenance-aware AI assistance and verification of externally circulated health claims; and (4) access histories, time-bounded permissions, and immediate access revocation. These capabilities are intended to make the boundaries between personal information, shared information, and AI-mediated interpretations understandable and actionable to the people involved. An underlying standardized profile representation supports the integration of heterogeneous data sources and the generation of consistent views across actors.

Before involving participants, we evaluated the system's core technical workflows using a synthetic corpus of 1,000 care circle configurations to assess data visibility, information-processing workflows, and policy-consistency mechanisms across diverse scenarios. 

Following this technical evaluation, we conducted a controlled laboratory study involving 13 subject\textendash care-actor dyads (26 participants) to investigate the immediate usability, cognitive legibility, and relational implications of \textit{JEEVHITAA}'s core interactions. Subjects wore a smartwatch for 24 hours before their laboratory session to provide authentic personal telemetry rather than relying exclusively on fabricated interface examples. During the task-based walkthrough, dyads engaged with three scenarios: setting up role-specific sharing settings, using the AI Verification Engine to investigate a circulating health claim, and inspecting the access history to revoke a care actor's access. The tasks were followed by dyadic semi-structured interviews and usability assessments.

Our laboratory evaluation was guided by three interconnected research areas:

\begin{enumerate}
    \item \textbf{RQ1: Usability and Acceptability of Collective Health Interactions.} How do subjects and care actors understand and use \textit{JEEVHITAA}'s shared dashboards and role-aware information-sharing controls?
    \item \textbf{RQ2: AI-Supported Sensemaking and Information Provenance.} How do participants interpret and use provenance indicators, source citations, and confidence information when examining wearable data and evaluating externally circulated health claims?
    \item \textbf{RQ3: Privacy, Transparency, and Relational Coordination.} How do participants perceive access transparency, revocability, and AI-mediated health briefings in relation to personal agency, surveillance concerns, and interpersonal care practices?
\end{enumerate}

We examine these research questions through task-based interactions, usability assessments, comprehension activities, and qualitative accounts, aiming to reveal how people understand, interpret, and negotiate interaction mechanisms.

This paper makes three contributions. First, it presents \textit{the design and implementation of an integrated mobile HCAI ecosystem that operationalizes collective health-information interactions} through role-aware dashboards, granular disclosure controls, provenance-supported AI assistance, and auditable access management. Second, it describes a \textit{context-informed synthetic evaluation strategy for exercising data visibility, information processing, and policy consistency mechanisms} across diverse care circle configurations prior to human evaluation. Third, it provides \textit{empirical evidence} from a controlled dyadic laboratory study that characterizes how subjects and care actors understand, evaluate, and negotiate these interaction mechanisms, including tensions between transparency and surveillance, AI mediation and interpersonal responsibility, and automated information presentation and user judgment. Together, these contributions provide \textit{design and evaluation insights for developing interactive health technologies that accommodate the relational nature of everyday health and wellbeing management.}
\section{Related Work}

\textit{JEEVHITAA} draws on research at the intersection of collective health informatics, privacy and information governance, and human-AI interaction. These areas provide the foundations for understanding how health information is shared, interpreted, and acted upon when health management involves multiple people. \textbf{Specifically, \textit{JEEVHITAA} addresses these three dimensions by pairing role-conditioned care-circle views with granular, time-bounded, and auditable disclosure controls, alongside provenance-grounded AI summaries and claim verification.} The following review examines these three strands of research and identifies the gap that motivates an integrated approach to collective health information management.

\subsection{Collective Health Informatics and Collaborative Interaction}

Personal Informatics (PI) has traditionally conceptualized health tracking as an individual process of collecting, integrating, reflecting on, and acting upon personal data. Li et al.'s~\cite{10.1145/1753326.1753409} stage-based model provides a foundational account of these activities and the barriers individuals encounter when engaging with self-tracking technologies. However, the assumption that the person generating health data is also responsible for interpreting and acting upon it does not adequately represent situations in which health management involves multiple people. Research on family informatics extends this perspective by examining how families collectively track and manage health, including how tracking responsibilities and information needs are distributed across family members~\cite{10.1145/2998181.2998362,10.1093/jamia/ocac049}. Research on personal informatics in interpersonal contexts further demonstrates how tracking data becomes embedded in wider social relationships and support networks~\cite{10.1145/3274396}. Similarly, research on caring through data demonstrates how personal health information becomes embedded in social and emotional caregiving practices, in which family caregivers use and interpret data to support another person's health~\cite{10.1145/2998181.2998303}. Research on designing for co-responsibility also demonstrates that health-related behavior change can depend on the coordinated involvement of patients, partners, and healthcare professionals rather than the individual alone~\cite{10.1145/3357236.3395469}.

This shift from individual to collaborative health management has also informed research on systems that support coordination among multiple stakeholders. Amir et al.~\cite{10.1145/2702123.2702320} examine how computational systems can support the transition from care plans to coordinated teamwork, highlighting the importance of shared understanding and coordination in complex healthcare activities. Research on shared health informatics and family informatics similarly emphasizes the need to integrate perspectives and support information exchange across people involved in care~\cite{10.2196/10830,10.1093/jamia/ocac049}. Studies of family resilience in care coordination further demonstrate that families operate as adaptive systems in which responsibilities and support arrangements evolve in response to changing circumstances~\cite{10.1145/3686998}. Research on caregivers as boundary actors also highlights the intermediary roles family members may take between patients, healthcare professionals, and wider support networks~\cite{10.1145/3579545}. In older-adult care, work on caregivers' multifaceted roles and informal coordination under uncertainty further illustrates that care arrangements involve evolving responsibilities and information needs~\cite{10.1145/3613904.3642728,10.1145/3817013}. Research on shared understanding in children's behavioral health likewise highlights the importance of maintaining common interpretations across stakeholders involved in care~\cite{10.1145/3449095}. Together, these studies establish that collaborative health management involves more than distributing information: it also requires people to interpret information, establish shared understanding, and coordinate their respective responsibilities.

Recent systems have begun to examine how interactive technologies can support collective health practices across different care settings. FamilyBloom investigates family-centered health tracking through shared smartwatch and home-display interactions, highlighting the opportunities and tensions that emerge when tracking spans household routines and wider care networks~\cite{10.1145/3772318.3791277,10.1145/3832000}. Family In The Loop brings AI-supported information synthesis into institutional dementia care by providing tailored information views for family members and care staff~\cite{10.1145/3757599}. Recent research has also examined collective health informatics beyond formally organized clinical teams. Srinivasan et al.~\cite{srinivasan2026unpackingpersonalhealthinformatics} examine how personal tracking technologies are appropriated within broader social arrangements, highlighting the multiple participants and interdependent activities involved in the use of health informatics. These works demonstrate the value of designing for collective participation, while also raising questions about how different participants understand health information, how responsibilities are negotiated, and how individuals retain agency when their information is shared. These questions motivate further investigation into systems that support collective health management across diverse care relationships.

\subsection{Privacy, Transparency, and Control in Health Information Sharing}

Sharing personal health information introduces questions that extend beyond conventional access permissions. Health measurements can reveal sensitive aspects of an individual's physical condition, routines, and everyday behavior. Consequently, the appropriateness of disclosure depends not only on who receives the information but also on the context, purpose, and expectations surrounding its use. Contextual integrity frames privacy in terms of the appropriateness of information flows in relation to actors, information attributes, and transmission principles, rather than simply the presence or absence of access~\cite{Nissenbaum_2004}. Badillo-Urquiola et al.~\cite{10.1145/3272973.3273012} emphasize the importance of critically engaging with contextual privacy theories when designing interactive systems, highlighting that privacy expectations are shaped by the circumstances and relationships in which information flows occur. Research on the non-static nature of patient consent similarly demonstrates that privacy expectations can change with health circumstances, technology use, and relationships with information recipients~\cite{10.1145/2441776.2441838}. In caregiving contexts, these considerations become particularly important because individuals may depend on others for assistance while also wishing to retain control over their personal information.

Prior work on access control for home data sharing shows that users' preferred sharing policies can be multidimensional and that their mental models may not align with the controls provided by digital systems~\cite{10.1145/1753326.1753421}. Research on mobile sensing has further explored dynamic privacy controls that allow users to inspect collected data and adjust consent according to contextual factors such as time and place~\cite{10.1145/3613904.3642815}. These studies demonstrate that usable information governance requires more than a fixed permission decision. In health contexts, research on older adults' experiences with IoT health monitoring has also identified concerns about unauthorized access, confusing data-sharing policies, and the difficulty of navigating complex security settings~\cite{10.1145/3706598.3714019}. These concerns are particularly relevant when designing systems intended to support older adults and the people involved in their care.

Prior research has also demonstrated that health technologies can redistribute responsibilities and introduce additional forms of care work. Kaziunas et al.~\cite{10.1145/2998181.2998303} examine the data needs of family caregivers, illustrating how access to and interpretation of health information become intertwined with caregiving practices. Chen et al.~\cite{10.1145/2441776.2441789} similarly highlight the importance of recognizing caregivers' own needs and the competing responsibilities involved in providing support. These findings suggest that enabling access to health information should not automatically be equated with improving collaboration. The circumstances in which information is shared, the expectations attached to that sharing, and the distribution of responsibilities among participants also shape how health information is used.

For interactive health systems, these concerns raise questions about the legibility of information-sharing arrangements and the consequences of granting, modifying, or withdrawing access. A permission mechanism may be technically correct but remain ineffective if users cannot understand what information is exposed, who can access it, or how those conditions can be changed. Similarly, transparency mechanisms such as access histories may support accountability while simultaneously making interpersonal monitoring more visible. The challenge is therefore not simply to maximize transparency or restrict disclosure, but to support comprehensible information flows and meaningful individual control. Although prior research establishes the importance of contextual privacy, usable access control, and caregiver needs, these concerns remain important to consider alongside the collaborative interpretation and use of health information within shared digital environments.

\subsection{Human-AI Interaction and Health Information Sensemaking}

The increasing availability of wearable health data has expanded the need for systems that help users interpret measurements rather than merely display them. Research on self-tracking metrics demonstrates that people interpret wearable measures in relation to their own purposes, activities, and changing circumstances; metrics are therefore not necessarily self-explanatory indicators of health or well-being~\cite{10.1145/3706598.3713650}. Large Language Models (LLMs) offer opportunities to translate heterogeneous health information into more accessible summaries and explanations. Recent systems have explored LLM-supported interpretation and recommendation using wearable and passive sensing data, including personalized sleep-health support and open-ended sensemaking across multiple sensing sources~\cite{10.1145/3706598.3713852,10.1145/3749474}. However, the ability to generate understandable language does not necessarily establish the factual reliability or appropriateness of the resulting information. This distinction is particularly important when health information is consumed by multiple people who may have different levels of health literacy and different responsibilities for responding to it.

Research on human-AI interaction provides important guidance for designing systems that support user agency rather than encourage uncritical reliance on automated outputs. Amershi et al.~\cite{10.1145/3290605.3300233} establish interaction guidelines that emphasize communicating system capabilities, making uncertainty understandable, and supporting users in evaluating AI-generated information. Empirical work by Ahn et al.~\cite{10.1145/3613904.3642780} further examines the relationship between model interpretability, outcome feedback, and user trust, demonstrating why explanations and feedback should be considered in relation to how people evaluate AI systems. In medical contexts, Lim et al.~\cite{10.1145/3706598.3714058} investigate domain-aligned explanations for diagnostic AI, highlighting the importance of presenting information in representations that users can meaningfully interpret rather than relying exclusively on technically sophisticated explanations. Recent work on hybrid health explanations also examines how explanation formats can be made more understandable and useful for lay users of mobile health recommendations~\cite{10.1145/3772071}.

These concerns extend beyond interpreting model-generated explanations to evaluating the information supporting an AI response. In health-related interactions, users may encounter advice through messaging platforms, social networks, and informal communication channels, where the credibility of a claim is not always apparent. Research on consumer health information quality distinguishes related but non-identical concepts of information quality, credibility, and trust, underscoring the importance of specifying what users are being asked to evaluate~\cite{10.1145/3576840.3578331}. Systems that generate assessments without exposing their supporting evidence risk replacing one source of uncertainty with another. Recent work on augmenting peer health posts with scientific evidence and personal anecdotes illustrates how evidence from different sources can be brought together to help users reason about health-related uncertainty~\cite{10.1145/3772318.3791089}. Research on making mental health reports more accessible through LLM-based approaches also demonstrates the potential of language models to translate complex clinical information into more comprehensible representations for non-expert audiences~\cite{10.1145/3706599.3720208}. However, accessibility must be accompanied by means for users to inspect the origin and supporting evidence of generated information.

Taken together, this research points to the importance of AI-supported health sensemaking that combines accessible representations with transparency about the information underlying generated outputs. This is particularly relevant in collective health management, where different participants may interpret the same measurement, explanation, or recommendation differently. Supporting such interactions requires attention not only to the comprehensibility of AI-generated information but also to the extent to which its supporting data and evidence can be examined.

\subsection{Research Gap and Positioning}

The literature establishes the social and collaborative nature of health information management, the importance of contextual privacy and individual control, and the need for human-centered mechanisms that make AI-generated information understandable and inspectable. Recent systems have begun to bring some of these concerns together: FamilyBloom examines shared family health tracking across devices, while Family In The Loop uses AI-supported information synthesis to connect families and care staff in institutional dementia care. Other systems investigate LLM-supported interpretation of wearable or passive sensing data. These contributions demonstrate important progress, but address distinct settings and interaction priorities. An opportunity remains to examine how subject-governed information sharing, privacy-aware access, and evidence-oriented AI sensemaking can be integrated within a shared environment involving diverse care relationships.

\textit{JEEVHITAA} addresses this opportunity through an integrated health ecosystem that brings together shared health information, privacy-aware access to information, and AI-supported health sensemaking. The focus is not on treating any of these capabilities in isolation, nor on claiming that collaborative health or AI-supported care systems have not been developed before. Instead, it is on examining their interaction in a system that emphasizes role-conditioned views, metric-level and time-bounded disclosure, auditable information flows, and AI-supported summaries and verification. In doing so, the study investigates how an integrated approach can support the interpretation, sharing, and use of personal health information while attending to the agency and differing perspectives of those involved in care.
\section{System Design and Implementation}

The design of \textit{JEEVHITAA} was informed by prior qualitative work by Srinivasan et al.~\cite{srinivasan2026unpackingpersonalhealthinformatics}, which identified information sharing as a central requirement in personal health informatics involving family members and other support actors. Their work, incorporating surveys, semi-structured interviews, stakeholder interviews, and co-design activities, highlighted the need to establish what health information should be shared, with whom, and under what conditions. These findings informed a focused set of system requirements: granular controls over individual health metrics, role-specific representations of health information, explicit consent and approval mechanisms, and AI-generated outputs accompanied by confidence and provenance information.

These requirements were translated into an integrated mobile application and supporting technical architecture. The design process used iterative prototyping to refine the organization of health information, sharing controls, and role-specific views. The resulting interaction specifications also informed the underlying data representation and software components, connecting the information presented through the application to structured profile entities, actor roles, and access permissions.

\subsection{System Architecture}

\begin{figure}[ht!]
    \centering
    \includegraphics[width=.99\linewidth]{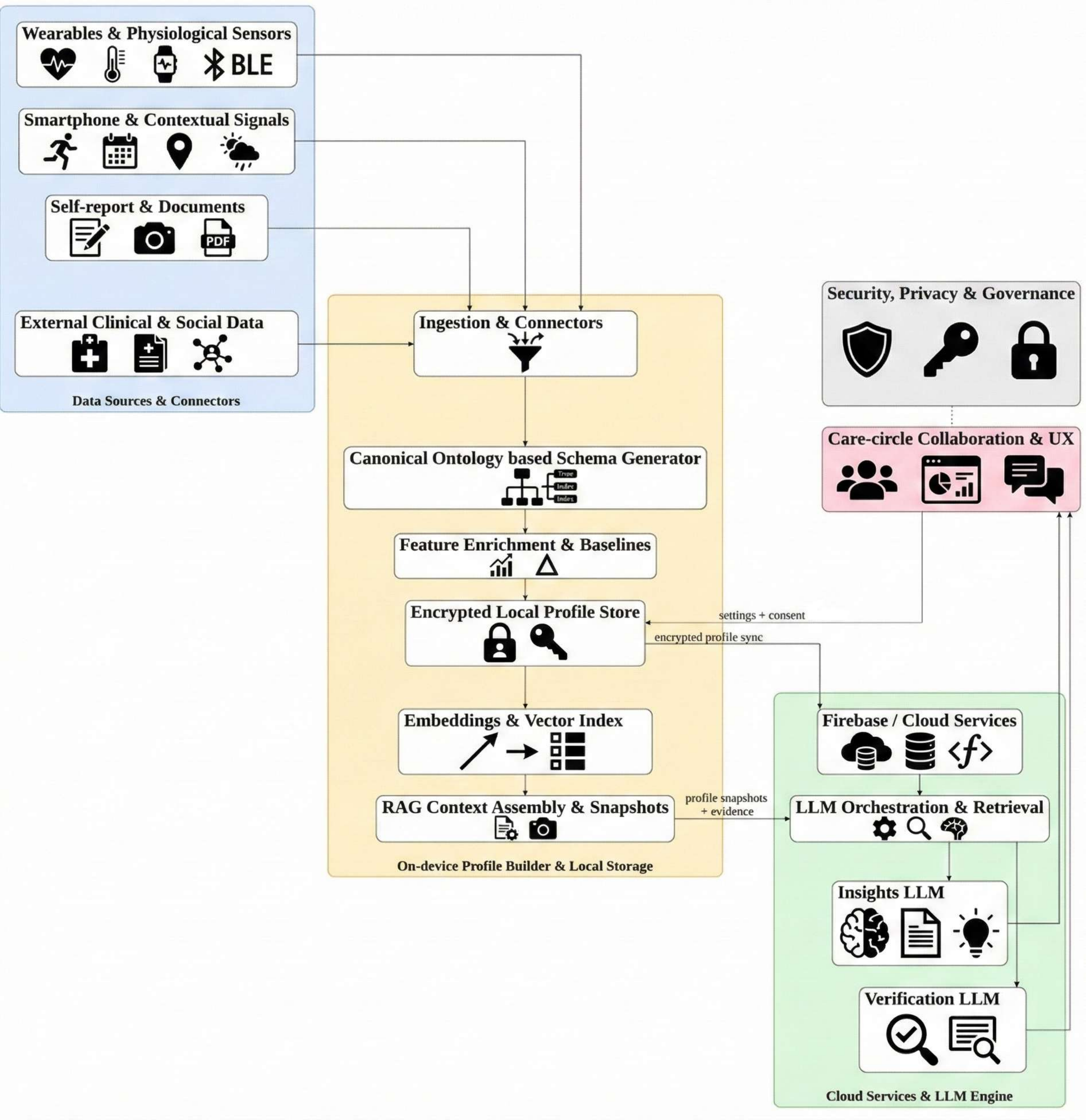}
    \caption{\label{fig:system-architecture} End-to-end system architecture of \textit{JEEVHITAA}.}
\end{figure}

\textit{JEEVHITAA} is organized as a mobile client supported by an ontology-driven profile builder, data-source connectors, local and cloud storage, and an AI orchestration layer. The architecture separates data acquisition and normalization from profile management, information sharing, and AI-assisted interpretation. Figure~\ref{fig:system-architecture} presents the end-to-end architecture and the principal data flows between these components.

The data pipeline begins with connectors that collect physiological, behavioral, and contextual information from supported sources. These include wearable sensors and BLE peripherals, smartphone sensors, platform health-data services, self-reported information, and optional clinical records. Incoming observations are represented as typed events with associated source and temporal information. The profile builder normalizes incoming values, maps them to the canonical profile representation, and derives selected higher-level features. The resulting profile entities and timeline events are persisted locally, while selected information is indexed for retrieval. Cloud synchronization is supported where enabled by the application.

The collaboration layer uses reusable, JSON-defined dashboard cards that reference fields in the profile schema. This allows health metrics, trends, alerts, and summaries to be rendered through different views without maintaining separate representations of the underlying data. Consent records and access-control rules determine which information is available to a given actor. The architecture therefore separates the canonical health profile from the role-specific views through which information is presented to other participants in a care circle.

The AI orchestration layer operates over structured profile information and retrieved evidence. It separates insight generation from information verification. The Insights Engine generates contextualized summaries and guidance using a compact profile snapshot and selected retrieved records. The Verification Engine evaluates health-related claims against supporting profile information and retrieved reference material, returning assessments with evidence references and confidence information. This separation supports the inspection of generated information without treating an LLM response as an authoritative clinical conclusion.

\subsection{Ontology and Profile Schema}

A central design decision was to represent personal health information through a structured, machine-readable ontology rather than relying on disconnected sensor-specific records or unstructured summaries. The ontology defines the entities and relationships used to represent an individual's health, circumstances, and participation in care-related activities. It provides a common representation through which information from different sensing sources, self-reports, and other inputs can be normalized and interpreted by the application.

The ontology specifies a \textit{Person} entity organized into three principal groups: \textit{IntrinsicFactors}, \textit{ExtrinsicFactors}, and \textit{CareAndInteractionFactors}. Intrinsic factors describe characteristics of the individual, including demographics, physiology, health conditions, and genetic predispositions. Extrinsic factors are the individual's surrounding context, including lifestyle behaviors, environmental conditions, socioeconomic circumstances, healthcare access, information sources, and access to technology. Care and interaction factors represent the structures and activities associated with collective support. These include actors and roles, care teams and plans, responsibility states, notification preferences, availability, consent and access policies, provenance, and audit information.

In the generated JSON representation, the subject's \texttt{person\_id} and the three principal groups are stored at the top level of the profile. The subject's participation in the care circle and the associated actor records are represented under \texttt{CareAndInteractionFactors.ActorsAndRoles.actors}. Each actor is assigned an actor identifier, a person identifier, and a role, together with relevant contact, visibility, delegation, authority, and availability attributes. The actor list can include the subject, family members, healthcare professionals, and other support actors. This structure allows the subject's profile to remain the canonical representation of their health information while enabling the application to construct different views according to each actor's role and permitted access.

The profile also represents care coordination and governance information through dedicated sections. Care teams and care plans describe team membership, goals, and notification types for self and care actors. Responsibility and state fields capture task ownership and transitions, while notification and availability fields describe communication preferences and actor availability. Policy, consent, provenance, and audit sections provide structured representations of information-sharing conditions and relevant records. The profile schema includes an administrative schema-version field to identify the representation's version. The complete field-level ontology and associated JSON structure are provided in the supplementary material.

\begin{figure}[ht!]
    \centering
     \includegraphics[width=.99\linewidth]{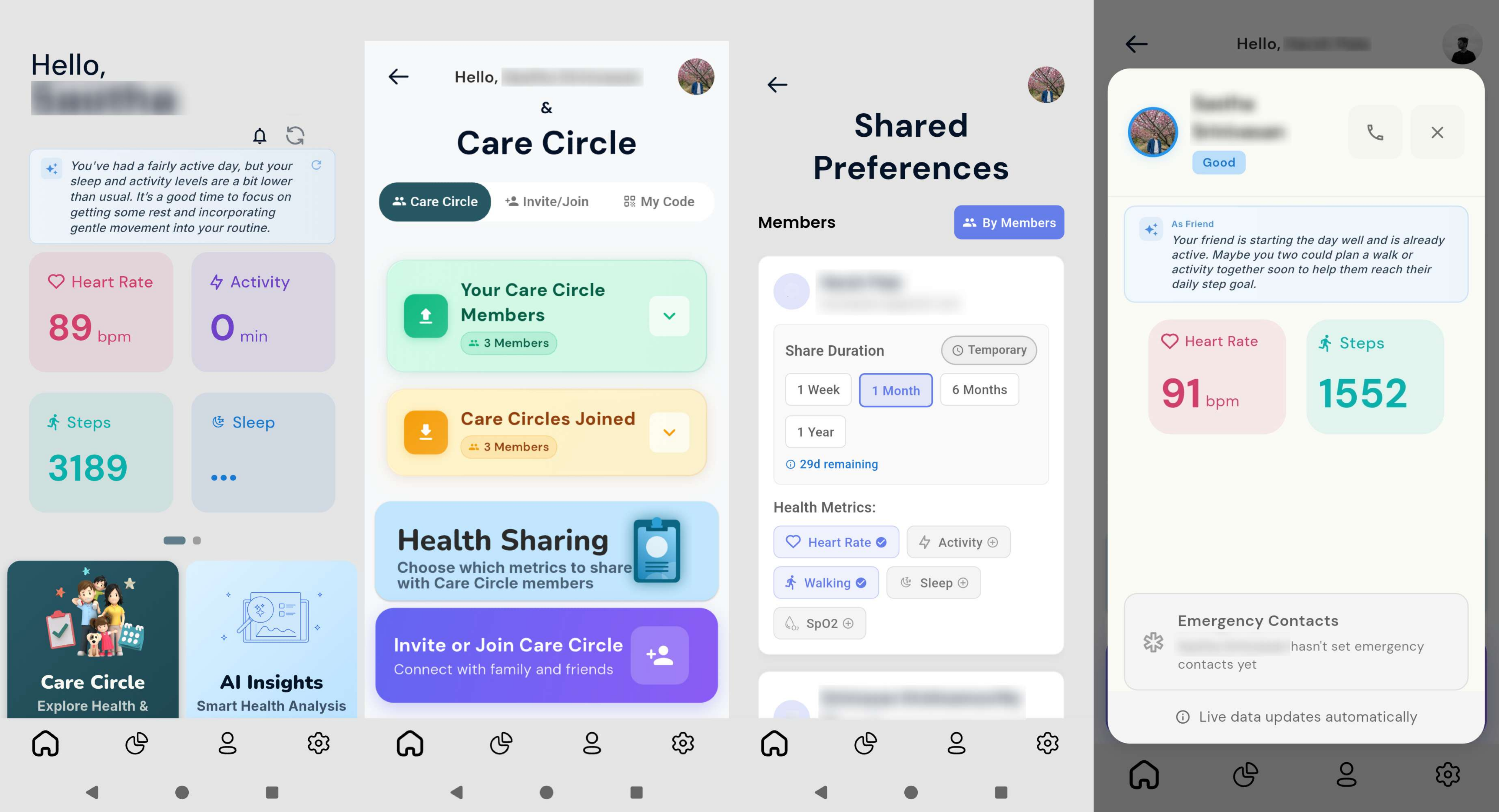}
     \caption{\label{fig:prof_flow} Android Interface - Profile, Sharing Controls Flow, Care Circles, and Shared Summary.}
\end{figure}

\begin{figure}[ht]
    \centering
     \includegraphics[width=.99\linewidth]{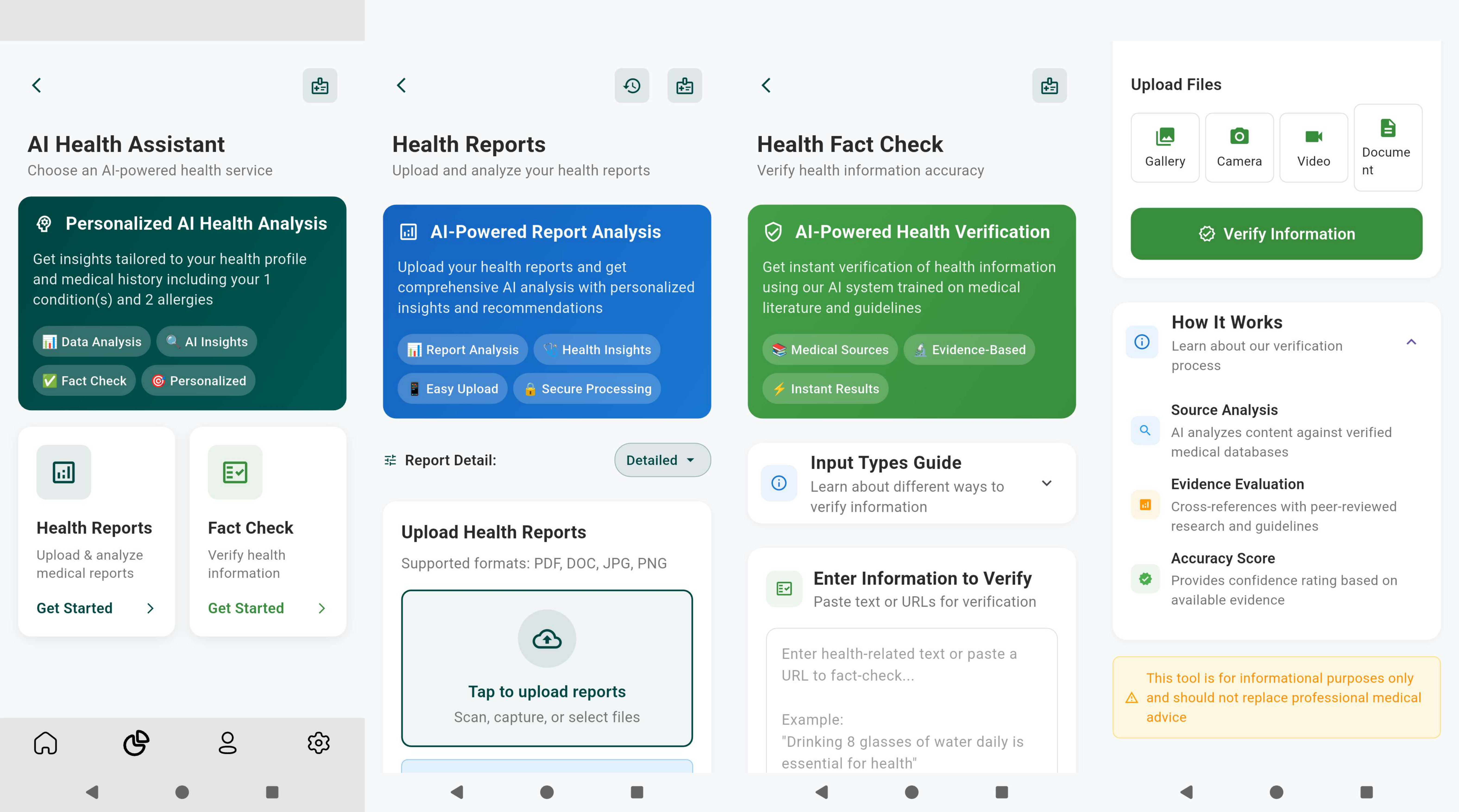}
     \caption{\label{fig:ins_ver} Android Interface - AI Assistance and Health Information Verification Engine.}
\end{figure}

\begin{figure}[ht!]
    \centering
     \includegraphics[width=.99\linewidth]{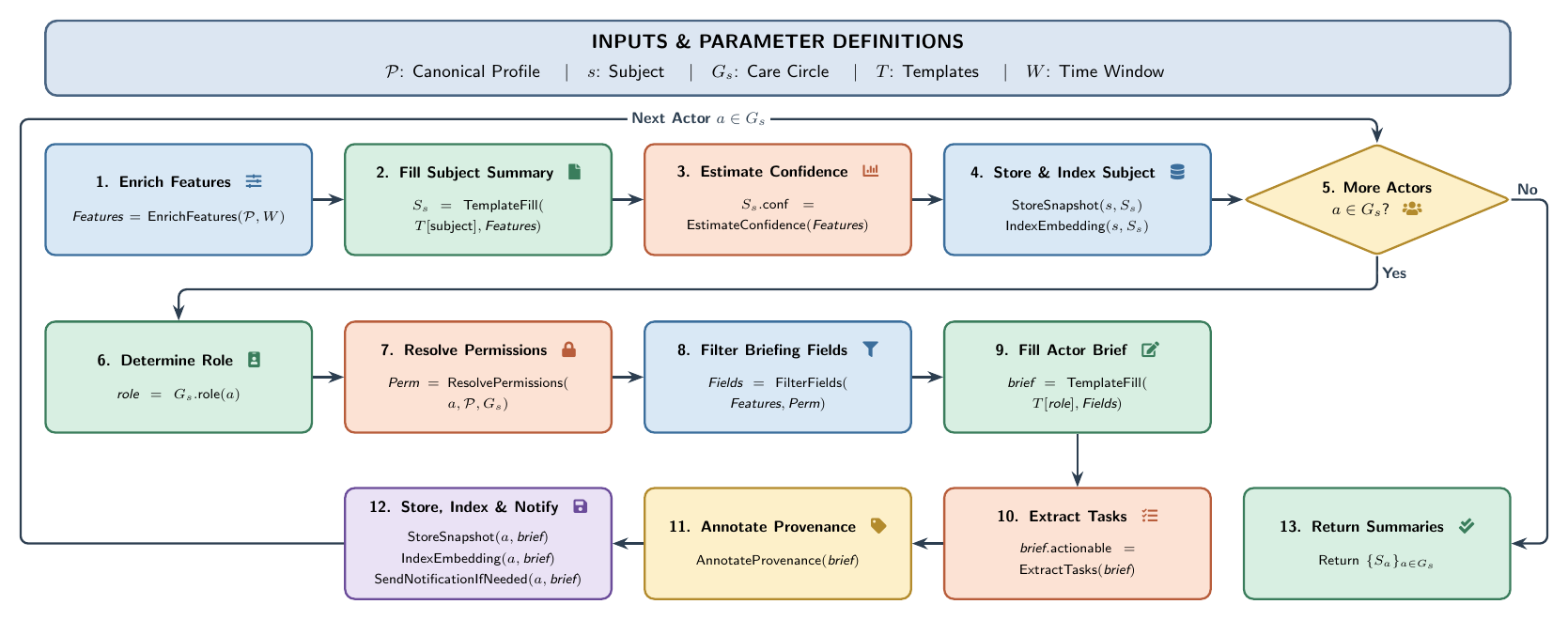}
     \caption{\label{fig:sum_algo} Algorithmic flow diagram of Summarizer.}
\end{figure}

\begin{figure}[ht!]
    \centering
     \includegraphics[width=.99\linewidth]{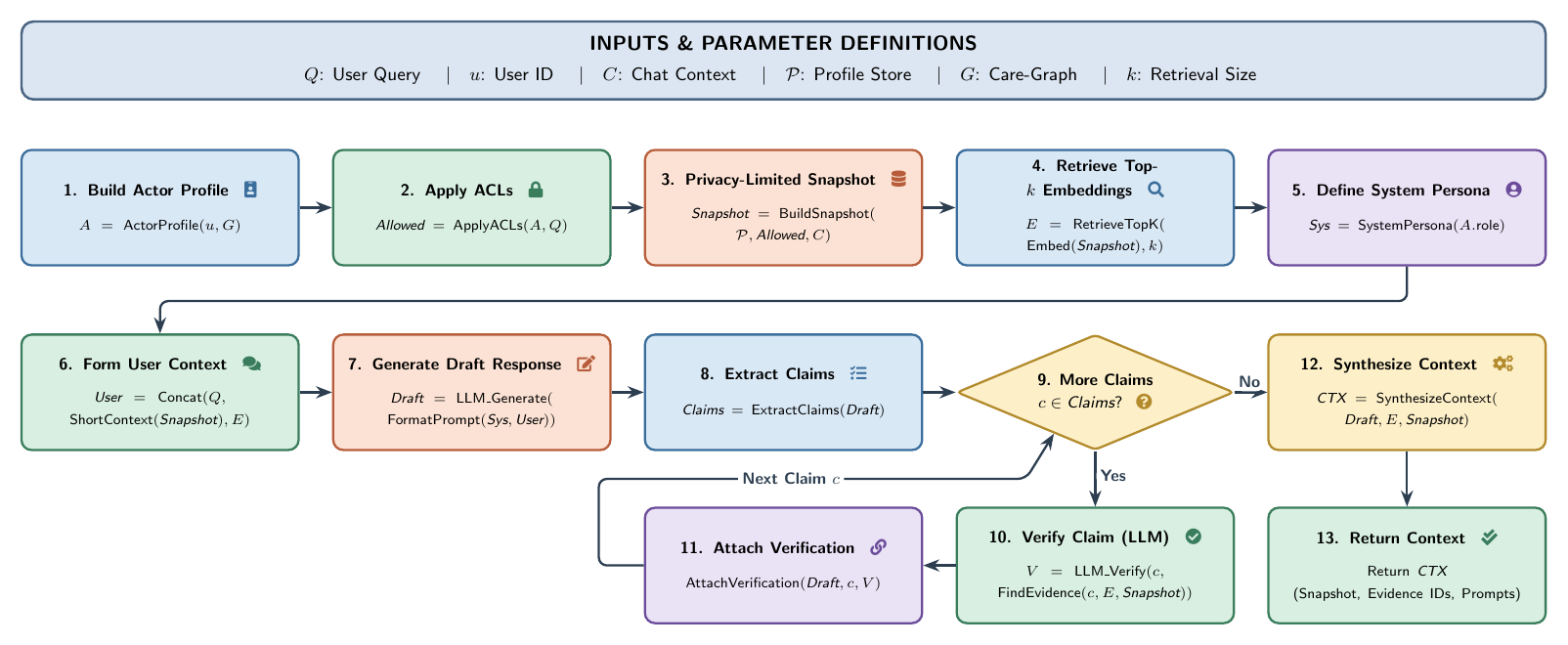}
     \caption{\label{fig:verf_algo} Algorithmic flow diagram of Verifer.}
\end{figure}

\subsection{Mobile Application and AI Implementation}

The initial implementation targets Android and is built using Flutter and Dart. The application separates presentation, domain, data, and service components, allowing interface elements and application logic to remain distinct from persistence and platform-specific integrations. Flutter provides application screens and reusable interface components, while native Android integrations support device and sensor access, background operations, notifications, and secure storage. Supported health data integrations include BLE-based peripherals and Android health data services such as Google Fit and Health Connect.

The application brings together profile dashboards, care circle views, sharing settings, and conversational AI interfaces (presented in FIgure~\ref{fig:prof_flow} \& ~\ref{fig:ins_ver}). Dashboard components are defined using reusable JSON card templates that reference ontology fields. This enables the same underlying profile information to be displayed in different configurations while applying the appropriate role and consent constraints. Users can also configure application preferences, including activity goals, step goals, daily time boundaries, and the models used for insight generation and verification.

The profile builder acts as the data transformation layer between incoming observations and the canonical profile. It performs ingestion, unit normalization, selected aggregation and feature derivation, and ontology mapping. Normalized profile records and timeline events are stored locally, with selected profile items and summaries converted into embeddings for retrieval. The vector index supports the construction of compact contextual windows for AI interactions, avoiding the need to include the entire profile history in each model request. Firebase services support authentication, profile persistence, synchronization, and server-side functions where required.

The AI functionality is implemented as two complementary components. The \textit{Insights Engine} (algorithm presented in Figure~\ref{fig:sum_algo} and Appendix~\ref{appendix:algo}) generates health summaries and contextual guidance using structured profile information, role-specific prompts, and retrieved evidence. For each interaction, the orchestration layer constructs a concise profile snapshot and retrieves a limited set of relevant profile entries or guideline excerpts. The model is instructed to produce structured outputs that include summaries, observations, recommendations, confidence levels, and references to supporting data. These outputs are subsequently rendered as human-readable content in the application.

The \textit{Verification Engine} (algorithm presented in Figure~\ref{fig:verf_algo} and Appendix~\ref{appendix:algo}) addresses a related but distinct task: assessing health-related information supplied by users, including claims encountered through messages, documents, links, or other external sources. Claims identified for assessment are evaluated against retrieved profile information and available reference material. The verifier returns assessments with evidence references and confidence information that can be inspected through the interface. The two components therefore serve different purposes: the Insights Engine contextualizes information about the subject, whereas the Verification Engine assesses the evidential support for an external health-related claim. Both are intended to assist interpretation and information assessment rather than replace professional medical judgment.

\subsection{Synthetic Care Circle Data Generation}

Synthetic care circle data were generated to provide varied and contextually structured profiles for system testing. The generation pipeline combines a context-generation runner, a profile-population module, the ontology specification, and reference datasets describing Indian health, socioeconomic, and environmental characteristics. The context-generation process combines health and life-stage themes, family structures, socioeconomic tiers, geographic regions, and care modalities. Compatibility rules are applied to selected combinations before contexts are sampled and passed to the profile generator.

For each context, the generator creates a subject profile and populates its ontology fields programmatically using predefined rules, value ranges, and probabilistic sampling. The reference datasets provide parameters for selected health and contextual attributes. These include age- and urbanicity-related disease prevalence, selected comorbidity relationships, socioeconomic and insurance distributions, healthcare access characteristics, and regional environmental conditions. For fields without a corresponding reference distribution, values are sampled from predefined ranges or categorical value pools. The generation process therefore combines context-specific reference parameters with scripted population rules; it does not treat the generated records as observations of actual individuals or as a representative sample of the population.

The generated subject profile is accompanied by an actor list constructed according to the scenario and the subject's characteristics. The actor-generation procedure includes the subject and may also include family members, healthcare professionals, and external support actors. The implementation limits each generated actor list to five entries, including the subject. Care teams and care plans are then generated using the available actors. A subsequent consistency step updates references to actor identifiers across relevant care-related fields. Structural post-processing also applies selected transformations and redactions to profile fields, including reduced birth-date precision and masking of selected location and insurance information.

During generation, a temporary top-level \texttt{metadata} object recorded information such as the care circle identifier, context description, health theme, family structure, socioeconomic category, region, care modality, generation timestamp, generator version, and actor count. This information was used during the generation process to track scenario assignments and examine consistency across the synthetic records. It was not part of the ontology or the subject's health profile. A separate preprocessing step removed the top-level metadata before the cleaned JSON profiles were used for testing. Accordingly, the test profile contains the subject's \texttt{person\_id} and ontology-organized profile groups, with actor and care-related information nested within \texttt{CareAndInteractionFactors}, but no generation metadata.

The synthetic dataset provided controlled inputs for exercising the system across varied health and care contexts, including differences in profile attributes, health conditions, socioeconomic circumstances, and care circle composition. It was used for functional and technical testing rather than epidemiological inference or claims about the prevalence of conditions in the Indian population.

\subsection{Laboratory Study Configuration}

For the controlled laboratory evaluation, the implemented system was configured around subject\textendash care-actor dyads. This configuration provided a bounded setting in which participants could interact with the subject's health information through distinct roles and access permissions. The dyadic configuration was specific to the laboratory study and did not impose a corresponding restriction on the system's intended support for larger care circles. Participant recruitment, study tasks, session procedures, and evaluation measures are described in the Methods section.
\section{Methods}
\label{sec:methods}

We conducted a controlled, mixed-methods laboratory study to examine the usability and cognitive legibility of \textit{JEEVHITAA}'s collective health interaction mechanisms, after approval from our institutional review board (IRB). The evaluation focused on three related areas: (1) how subjects and care actors understood shared health dashboards and role-based information access; (2) how participants interpreted AI-generated health information, provenance indicators, confidence scores, and supporting citations; and (3) how participants perceived privacy controls, access transparency, and the potential for system-mediated interactions to affect interpersonal friction. This guided evaluation had the following associated design hypotheses:

\begin{itemize}
    \item \textit{H1.1:} Participants will be able to identify the information visible to their care circle counterpart and locate the corresponding sharing and visibility controls during guided interactions.
    \item \textit{H1.2:} Subjects will perceive metric-level, subject-controlled, and time-bounded sharing as supporting disclosure decisions that better reflect their relationship-specific care responsibilities and privacy preferences.
    \item \textit{H2.1:} Participants will use the Verification Engine and its inspectable citations to examine the evidence supporting assessments of external health claims.
    \item \textit{H2.2:} Participants will interpret provenance indicators and confidence scores as supporting cues for evaluating information, while calibrating their reliance on AI outputs against source credibility and personal context.
    \item \textit{H3.1:} Participants will perceive system-generated health briefings as a potentially less confrontational means of initiating care conversations than repeated person-to-person reminders.
    \item \textit{H3.2:} Participants will perceive explicit access histories and revocation controls as mechanisms for retaining awareness and control over disclosure, while recognizing that information visibility may also create surveillance concerns or implicit pressure to share.
\end{itemize}

The study combined task-based interaction, dyadic semi-structured interviews, individual usability questionnaires, and verbal comprehension prompts. This configuration enabled examination of both participants' immediate interactions with the system and their interpretations of its mechanisms within an existing care relationship.

\subsection{Study Design and Laboratory Procedure}

The study used subject\textendash care-actor dyads as the unit of interaction. Each dyad consisted of a subject, whose health information was represented in the application, and a care actor who participated in reviewing and interpreting the shared information. The dyadic configuration was selected to control for interactions between participants with different relationships to the same health data. It was an evaluation configuration rather than a limitation on \textit{JEEVHITAA}'s intended support for larger care circles.

To reduce variation due to differences in wearable hardware and device pairing configurations, the study used a standardized device setup. Each subject wore one of the two available Samsung Galaxy Watch Ultra (A/B) for 24 hours before the laboratory session. This provided recent personal wearable telemetry for review rather than relying exclusively on dummy or illustrative data. A dedicated Samsung Galaxy Z Flip 6 smartphone was used for the in-lab application walkthrough; the paired smartphone was retained by the principal investigator until the evaluation session. The application account was prepared in advance for both the primary participant and a care-circle member, allowing the walkthrough to focus on initial setup, information visibility, sharing controls, and role-specific interactions rather than on account configuration alone.

Care actors attended either in person or via video call. During the interactive tasks, the subject used the application to inspect the health information and sharing settings, while the care actor observed the corresponding display and participated in the discussion. Each completed laboratory session lasted approximately 90\textendash 120 minutes. The protocol allocated approximately five minutes for the initial briefing, 20 minutes for the interactive walkthrough, up to 50 minutes for the dyadic interview, and 15 minutes for the questionnaires and comprehension assessment. Actual session duration varied with the pace of interaction and the time required for discussion.

The interactive walkthrough comprised three tasks:

\begin{enumerate}
    \item \textbf{Wearable dashboard and care-circle sharing.} Participants reviewed the subject's 24-hour wearable summary and inspected the care-circle dashboard. They examined which health metrics were visible to the care actor, identified the relevant sharing controls, and considered which information should remain private or be made available to the other participant. The task focused on whether participants could understand the relationships among the subject's profile, the care actor's role, and the information presented in the shared view.

    \item \textbf{AI verification and provenance inspection.} Participants were requested to verify a health-related claim encountered in informal digital communication. They used the AI Verification Engine to assess the claim, inspect the displayed confidence information, and review the supporting citations. The task focused on how participants interpreted the assessment, located and inspected its evidence, and evaluated the authority of the linked sources.

    \item \textbf{Access history and immediate revocation.} Participants considered a scenario in which the subject wished to restrict the care actor's access to specific health metrics. They located the access history, examined previous access information, and used the revocation control to withdraw the relevant access. Participants were also asked to explain their understanding of the revocation process.
\end{enumerate}

During the walkthrough, attention was directed to participants' navigation through the application, their understanding of visibility boundaries, their interpretation of provenance and confidence information, and their ability to locate and use the access-management controls. These interactions provided a concrete basis for the subsequent interview rather than relying exclusively on hypothetical descriptions of system use.

\subsection{Participants and Dyadic Composition}

The participants were recruited using purposive and snowball sampling to ensure inclusion of subjects of care who could attend the lab study component in person. The study was initially conducted with 16 subjects; however, 3 of them dropped out at various stages for personal reasons. The final data collected for the study involved 13 subject\textendash care-actor dyads, comprising 26 participants. The participant information records contained demographic information for each subject and the corresponding care actor, including age, gender, relationship, and the subject's educational background, technology proficiency, mobile ecosystem, prior health app use, and frequency of contact with the care actor.

The 13 subjects were aged 18\textendash 37 years ($M=26.3$, $SD=6.2$); nine identified as female and four as male. Their care actors were aged 19\textendash 56 years ($M=30.1$, $SD=11.2$), comprising seven females and six males. Across the full sample, participants were aged 18\textendash 56 years. The care relationships included spouses, parents, siblings, cousins, friends, and roommates. This composition enabled examination of the same interaction mechanisms across family and non-family relationships.

Among the subjects, eight had completed a Master's degree, two had completed a Bachelor's degree or diploma, and three were pursuing their undergraduate degree. Twelve subjects reported advanced technology proficiency, and one reported intermediate proficiency. Nine used Android devices, and four used iOS devices. Prior health app experience varied: three subjects had never used a health app, while the remainder reported experience ranging from occasional or one-time use to regular or daily use. Reported contact with the care person also varied: seven subjects reported daily contact, two reported contact several times a week, and four reported contact approximately once a month. Table~\ref{tab:lab-participants} presents the participant and dyadic characteristics recorded during screening.

\begin{table*}[ht!]
\centering
\scriptsize
\setlength{\tabcolsep}{3pt}
\resizebox{\textwidth}{!}{
\begin{tabular}{llllllll}
\toprule
\textbf{Dyad ID} &
\parbox[t]{1.2cm}{\raggedright \textbf{Subject}\\\textbf{(age/gender)}} &
\textbf{Education} &
\parbox[t]{1.2cm}{\raggedright \textbf{Technology}\\\textbf{proficiency}} &
\textbf{Mobile OS} &
\textbf{Prior health-app use} &
\parbox[t]{2.4cm}{\raggedright \textbf{Care actor}\\\textbf{(relationship;}\\ \textbf{age/gender)}} & 
\parbox[t]{1.2cm}{\raggedright 
\textbf{Contact}\\\textbf{frequency}}\\
\midrule
JLSPWA1 & 19 / M & High school & Advanced & Android &
1\textendash 2 times/week; Google Fit, Strava &
Roommate; 19 / M &
Daily \\
JLSPWB1 & 19 / M & High school & Advanced & iOS &
\parbox[t]{3cm}{1--2 times/week; Apple Health, Apple Fitness, Samsung Health, Planfit} &
Close friend; 19 / M &
Daily \\
JLSPWA2 & 25 / F & Master's & Advanced & Android &
1\textendash 2 times/week; Samsung Health &
Brother; 20 / M &
Daily \\
JLSPWB2 & 25 / M & Bachelor's & Advanced & Android &
Never used a health app &
Close friend; 24 / F &
Daily \\
JLSPWB3 & 21 / F & Bachelor's & Advanced & Android &
Never used a health app &
Friend; 21 / F &
Monthly \\
JLSPWA4 & 25 / F & Master's & Advanced & Android &
Less than once/week; app not specified &
Friend; 23 / F &
Daily \\
JLSPWB4 & 34 / F & Master's & Advanced & iOS &
Less than once/week; Apple Health &
Husband; 35 / M &
Daily \\
JLSPWA5 & 18 / M & High school & Advanced & Android &
1\textendash 2 times/week; Google Fit &
Father; 44 / M &
Daily \\
JLSPWB5 & 28 / F & Master's & Advanced & Android &
Used once or twice; Healthify, Flo &
Mother; 56 / F &
Daily \\
JLSPWA6 & 28 / F & Master's & Advanced & iOS &
Never used a health app &
Sister; 32 / F &
Daily \\
JLSPWB6 & 35 / F & Master's & Advanced & Android &
\parbox[t]{3cm}{3--5 times/week; Samsung Health, Healthify Me, Cultfit}&
Husband; 37 / M &
Monthly \\
JLSPWA7 & 28 / F & Master's & Advanced & Android &
3\textendash 5 times/week; MyFitnessPal, Flo &
Sister; 26 / F &
Monthly \\
JLSPWB7 & 37 / F & Master's & Intermediate & iOS &
Daily; fitness application &
Cousin; 35 / F &
Daily \\
\bottomrule
\end{tabular}
}
\caption{\label{tab:lab-participants} Participant and dyadic characteristics ($N=26$, comprising 13 subjects and 13 care actors). Ages and genders for care actors were reported in the subject participant information form. M = male; F = female.}
\end{table*}

For participant-level identification in the transcripts and thematic analysis, each dyad identifier was extended with a role suffix: \texttt{S} denotes the subject and \texttt{C} denotes the care actor. Thus, while the participant table retains the dyad-level identifier (e.g., \texttt{JLSPWA1}), the corresponding individual participants are identified as \texttt{JLSPWA1S} and \texttt{JLSPWA1C}, respectively. This convention was maintained throughout the coding process and the accompanying thematic analysis codebook.

\subsection{Interview Protocol and Qualitative Data Collection}

Following the interactive walkthrough, each dyad participated in a semi-structured interview with both the subject and the care actor. The interview guide (Appendix~\ref{appendix:interviewguide}) was organized into two broad parts: existing health-support practices and participants' evaluations of the system.

The first part established the relational and informational context in which the application was being evaluated. Participants were asked to describe recent instances in which they had sought or provided help with a health-related task, how responsibilities were distributed, and how they decided whom to involve. Questions also explored how participants encountered health advice through sources such as messaging platforms and online media, how they assessed the credibility of such information, and whether health-related reminders or monitoring had caused tension between them.

The second part examined participants' experiences of the system following the walkthrough. Questions were organized around the three research questions. For RQ1, participants discussed whether the information shown to the care actor matched their expectations, whether role-specific visibility boundaries were understandable, and whether the shared summary presented an appropriate level of detail. For RQ2, questions explored whether participants could identify the origin of wearable measurements, how they interpreted the Verification Engine's assessment, and how they evaluated confidence indicators and citations. For RQ3, participants reflected on the possible effect of system-generated summaries on repeated reminders, their comfort with sharing health information, and the implications of being able to inspect access histories and revoke permissions.

The interview guide included questions directed separately to subjects and care actors where their perspectives differed. For example, subjects were asked about the experience of being monitored and controlling access to their information, while care actors were asked about the responsibilities and concerns associated with supporting another person's health. Follow-up questions were used to clarify participants' interpretations and to elicit concrete examples from their interactions with the application. The interviews therefore examined both the perceived usefulness of individual interface components and participants' interpretations of those components within the context of their relationship.

The 13 available dyadic interview recordings ranged from approximately 29.1 to 54.6 minutes ($M=41.1$ minutes, $SD=8.2$ minutes). These values refer to the originally recorded interview segments and not to the overall laboratory session duration, which also included the initial briefing, interactive tasks, and assessment activities.

\subsection{Usability and Comprehension Measures}

Three complementary instruments were used to capture participants' evaluations of the system and their understanding of its interaction mechanisms.

\subsubsection{Component Usability Matrix}

The Component Usability Matrix assessed four principal interface components: (1) the wearable telemetry dashboard, (2) the role-based sharing dashboard, (3) the AI Verification Engine and confidence indicators, and (4) the access logs and revocation controls. Participants rated each component across three dimensions: visual clarity, ease of interaction, and perceived trustworthiness of the information. Each dimension used a five-point Likert scale, with higher ratings indicating more favorable evaluations. The questionnaire (Table~\ref{appendix:componentusability}) was completed individually by subjects and care actors, yielding 26 responses.

\subsubsection{System Usability Scale}

Overall perceived usability was assessed using the standard ten-item System Usability Scale (SUS) (Appendix~\ref{appendix:sus}). Participants responded to each item on a five-point agreement scale, ranging from strongly disagree to strongly agree. SUS scores were calculated using the standard scoring procedure. The instrument was completed individually by both members of each dyad, producing 26 responses.

\subsubsection{Verbal Comprehension and Provenance Assessment}

A verbal comprehension assessment (Appendix~\ref{appendix:comprehension}) was administered at the end of the session to examine whether participants could interpret the system's provenance and privacy mechanisms in practical terms. The prompts covered three areas. First, participants were asked to identify the most recent access to their health information and distinguish wearable-derived data from manually entered data. Second, they were asked to locate citations that support a verification result and to explain how they would interpret a low-confidence assessment, with sources. Third, they were asked to explain what would happen when a time-bound permission expired and to identify how immediate revocation of access affected the care actor's view.

The assessment was designed to examine participants' explanations and demonstrated understanding, rather than solely their subjective confidence in the interface. Responses were considered alongside observations from the task walkthrough and accounts from the subsequent interview.

\subsection{Data Analysis}

The study used a combination of descriptive quantitative analysis and qualitative thematic analysis. The two forms of analysis addressed complementary aspects of the evaluation: questionnaire responses provided structured assessments of usability and interface components, while interview and comprehension data captured participants' interpretations, expectations, and relational concerns.

\subsubsection{Quantitative Analysis}

The SUS responses were scored using the standard ten-item procedure described above and summarised using descriptive statistics. Component Usability Matrix responses were examined descriptively across the four interface components and the three rating dimensions of visual clarity, ease of interaction, and information trustworthiness. Both instruments contained 26 participant responses, representing the subject and care-actor perspectives across the 13 dyads. The analysis retained the participant role associated with each response, allowing ratings to be examined in relation to the two positions within a dyad.

\subsubsection{Hybrid Deductive\textendash Inductive Thematic Analysis}

The qualitative material was analyzed using a hybrid deductive\textendash inductive thematic approach, drawing on the method described by Fereday and Muir-Cochrane~\cite{fereday2006demonstrating}. The initial deductive structure was informed by the study's three research questions and associated hypotheses. It provided an organizing framework for examining role comprehension and information sharing (RQ1), AI-supported sensemaking and provenance (RQ2), and privacy, transparency, and relational friction (RQ3). This initial structure was not treated as exhaustive; the analysis also retained patterns that extended or qualified the expectations established by the protocol.

The thematic analysis proceeded through an iterative process of familiarization, coding, code refinement, and theme development. The primary researcher read and re-read the manually verified interview transcripts to become familiar with participants' accounts and identify meaningful segments relevant to the research questions. An initial deductive coding structure was developed from the three research questions and associated hypotheses, capturing participants' experiences of role comprehension and information sharing (RQ1), AI-supported sensemaking and provenance (RQ2), and privacy, transparency, and relational friction (RQ3). These initial codes were applied to relevant segments of the transcripts. During subsequent engagement with the material, additional open codes were developed inductively to capture experiences and interpretations that were not adequately represented by the initial framework.

The first-order codes were iteratively examined, compared, and refined. Their definitions and analytical boundaries were clarified as recurring patterns, and distinctions became apparent across participants' accounts. Related first-order codes were then grouped into second-order subthemes, which were further organized into overarching themes. This process resulted in 35 first-order open codes, 23 second-order subthemes, and five overarching themes. Each code and subtheme was documented with an analytical definition, representative transcript evidence, participant identifiers, and its deductive [D] or inductive [I] classification, as appropriate. The resulting codebook provides a traceable account of how the initial research-question-informed categories were extended through engagement with the interview material.

The entire coding process was conducted by a single researcher, who was responsible for familiarization with the transcripts, developing and applying the initial coding framework, identifying inductive codes, refining code definitions, and organizing the resulting subthemes and overarching themes. No second independent coder was involved. Consequently, inter-coder agreement or inter-rater reliability statistics were not calculated. The coding process is therefore reported as an individual researcher's iterative qualitative analysis rather than as an independently replicated coding exercise. The accompanying appendix (Appendix~\ref{appendix:coding}) documents the resulting analytical structure, code definitions, representative evidence, and participant identifiers to provide transparency into the interpretation and organization of the qualitative material.

\begin{enumerate}
    \item \textbf{Collective Care Roles and Shared Health Management:} How participants understood care-circle membership, role-specific visibility, shared dashboards, metric-level disclosure, time-bounded access, and the responsibilities associated with supporting another person.

    \item \textbf{AI-Supported Sensemaking, Provenance, and Trust Calibration:} How participants identified data origins, inspected citations, interpreted confidence information, evaluated external health claims, and assessed the contextual relevance and actionability of AI-generated guidance.

    \item \textbf{Relational Friction, Privacy, and Surveillance Governance:} How participants interpreted access histories, disclosure boundaries, monitoring, and accountability, including the possibility that transparency could provide reassurance for some participants while creating feelings of surveillance or disclosure pressure for others.

    \item \textbf{Interaction Legibility and Technical Reliability:} Participants' experiences of navigating the application, understanding configuration settings, receiving explanations about data use, and accessing concise but sufficiently informative dashboards.

    \item \textbf{Personal Agency and Reflective Engagement:} How participants interpreted personal goal and notification configuration, and how application usage information could support reflection on their own health concerns and priorities.
\end{enumerate}

The first three themes included protocol-aligned deductive categories and inductive extensions, while the latter two captured additional patterns that were observed from participants' accounts. Examples included the importance of relationship-specific expectations for monitoring, the influence of notification frequency and summary tone on perceived interpersonal friction, requests for plain-language explanations of health metrics, and the use of application engagement information as a form of self-reflection.

The analysis retained participant identifiers and speaker-role distinctions throughout the codebook. Subject accounts and care-actor accounts were therefore not collapsed into a single undifferentiated participant perspective. This enabled examination of convergent and divergent interpretations within the dyads, including differences in how subjects and care actors understood visibility, responsibilities, reassurance, and privacy. The resulting thematic structure is documented in the accompanying codebook, which links each subtheme to its constituent first-order observations, participant identifiers, and relevant research questions.

\subsubsection{Integration of Data Sources}

The qualitative and quantitative materials were considered together to develop a more complete account of participants' experiences. Interview accounts were examined alongside the Component Usability Matrix and SUS responses to contextualize participants' perceptions of interface clarity, ease of interaction, and overall usability. Similarly, participants' explanations during the provenance assessment were considered alongside their interview accounts about source attribution, confidence indicators, citations, and access controls. Observations from the interactive tasks provided additional context for interpreting statements about navigation, information visibility, and the use of the revocation mechanism.

This integration was used to identify areas of convergence and divergence between participants' reported experiences, structured usability ratings, and demonstrated understanding. For example, a participant's favorable assessment of a sharing control was considered in relation to their ability to explain what information the care actor could access. Likewise, positive interpretations of confidence indicators were examined in the context of participants' descriptions of source inspection and independent verification. The analysis did not assume that favorable usability ratings necessarily indicated accurate comprehension or unqualified trust in AI-generated information.

\subsection{Transcription and Transcript Verification}

The dyadic interview recordings were initially transcribed using Google's Gemini Pro. All generated transcripts were subsequently checked manually against the corresponding audio recordings before being used for analysis. This review included checking the accuracy and completeness of participant statements, speaker attribution, relevant terminology, and contextual details. Errors or omissions identified during the review were corrected against the original audio. The manually checked transcripts, rather than the unverified machine-generated output, served as the qualitative material for the subsequent analysis. Participant identifiers were used to maintain the anonymity of subjects and care actors in the analytical records.
\section{Evaluation and Findings}
\label{sec:evaluations_and_findings}

This section reports two complementary evaluations of \textit{JEEVHITAA}: a system-level technical evaluation and a controlled laboratory evaluation involving 13 subject\textendash care-actor dyads (26 participants). The system-level evaluation examined the technical operation of the application suite, including role-conditioned summary generation and verification using a synthetic care-circle corpus, role-based information exposure, and compatibility with wearable data import across device ecosystems. The laboratory evaluation examined how participants interacted with the implemented interfaces and interpreted the information presented by the system. Its findings comprise descriptive usability ratings from the System Usability Scale (SUS) and Component Usability Matrix, observations of participants' interactions with the application, and qualitative findings from the 13 subject\textendash care-actor interviews. The qualitative analysis is organized around the five overarching themes and 23 second-order subthemes presented in the thematic analysis codebook.

These evaluations address different aspects of the system and should be interpreted separately. The system-level technical evaluation assesses generated content against role-specific reference information, examines source attribution and hallucination errors in verification, checks whether information exposure conforms to predefined access scopes, and assesses the availability of wearable data through the common ingestion layer. These tests provide evidence about technical performance under the evaluated synthetic profiles, access configurations, and device-integration paths; they do not establish clinical validity or the accuracy of the underlying wearable measurements. The laboratory study, in turn, provides evidence about participants' perceptions and experiences during controlled interaction with the application. It did not include an independently labeled benchmark for assessing the factual accuracy of generated summaries or verification responses, nor a controlled comparison condition to measure changes in interpersonal friction. Accordingly, questionnaire ratings of information trustworthiness are reported as participants' perceptions, and accounts of reduced reminders or social pressure are distinguished from observed changes in ongoing behavior. Similarly, task interactions are described through recorded observations and interview accounts rather than converted into task-completion percentages, as no standardized item-level task-scoring record was maintained.

\subsection{System-Level Technical Evaluation}

\subsubsection{Synthetic Care-Circle Corpus and Evaluation Configuration}

The technical evaluation was conducted using the full corpus of 1,000 synthetically generated care circles, each represented using the system's complete profile schema. The generated care circles had a median size of four members (IQR: 3\textendash 5), with each profile capturing the subject's health-related information, contextual factors, and relationships with actors involved in their care. The corpus was used to evaluate role-conditioned summary generation, verification, and information exposure across different care contexts.

Two versions of Gemma 4 were used in the initial language-model experiments. The evaluation covered the summary-generation component and the separate verification component. For summary generation, each test case was associated with a reference response constructed programmatically from the corresponding care-circle profile and the target actor's role and permitted information scope. This process established the information and actions expected to appear in a role-conditioned summary without relying on an LLM to generate the reference. It also made it possible to identify information that should not be disclosed to a particular actor. The generated summaries were assessed against these references to determine whether expected information was included and whether restricted information or words were exposed.

\subsubsection{Summary Generation and Verification Results}

The summary-generation component was evaluated using precision, recall, and F1-score. These measures captured the correspondence between the generated summaries and the expected role-specific information and actions. At the level of the expected information or action elements, correctly included elements contributed to true positives, omitted expected elements to false negatives, and unexpected or restricted information appearing in the summary to false positives. In particular, the evaluation treated the exposure of information or words that should have remained restricted as leakage rather than as valid summary content. The verification component was evaluated separately, focusing on source attribution and hallucination errors. Source attribution concerned whether generated information could be traced to the underlying profile data, while hallucination assessment concerned the presence of claims that were not supported by the available source information. The verifier's precision, recall, and F1-score reflect its performance on these verification dimensions. Table~\ref{tab:summary-verification-results} presents the reported results for the two components.

\begin{table}[ht!]
\centering
\begin{tabular}{lccc}
\hline
\textbf{Component} & \textbf{Precision} & \textbf{Recall} & \textbf{F1-score} \\
\hline
Summary generator & 0.94 & 0.89 & 0.91 \\
Verification component & 0.92 & 0.88 & 0.90 \\
\hline
\end{tabular}
\caption{\label{tab:summary-verification-results}Technical evaluation results for summary generation and verification.}
\end{table}

The results indicate the extent to which the generated content conformed to the expected role-specific information and whether the verification component identified source-attribution and hallucination-related issues. These metrics assess content correspondence and verification performance within the evaluated synthetic profiles; they should not be interpreted as evidence of clinical validity or as a substitute for clinical evaluation.

\subsubsection{Role-Based Information Exposure Checks}

In addition to evaluating generated content, role-based information exposure was examined against the predefined stakeholder views. The checks assessed whether each role received the information permitted by its configured access scope and whether information outside that scope was exposed. The Care Primary view passed all 15 expected-field checks, the Monitor view passed all seven, and the Nudge-only view passed all three. No additional fields were observed outside the predefined stakeholder views in these checks.

A separate automated constraint-checking process examined 120 edge-case configuration templates and flagged two configurations involving policy conflicts that required attention. These results are limited to the tested views and configuration templates; they do not establish that information exposure is impossible under all combinations of users, roles, and care contexts.

\subsubsection{Wearable Data Import Compatibility}

The application suite uses Google Health Connect as its common data-ingestion layer. Consequently, the wearable integration evaluation examined the extent to which data from different proprietary device ecosystems could be made available for import through this underlying layer. The reported pass rates therefore describe data-import compatibility across the tested ecosystems and integration paths, rather than the accuracy of the wearable sensors or the quality of the measurements they produce.

\begin{table}[ht!]
\centering
\begin{tabular}{lc}
\hline
\textbf{Device ecosystem or integration path} & \textbf{Import pass rate} \\
\hline
Google Health Connect (Fitbit/Pixel) & 100\% \\
Samsung native SDK & 98\% \\
Garmin cloud OAuth & 94\% \\
Mixed cloud-to-BLE integration & 88\% \\
Standalone ecosystems (Amazfit, boAt, Noise) & 89\textendash95\% \\
\hline
\end{tabular}
\caption{\label{tab:wearable-import-rates}Reported wearable data-import pass rates across device ecosystems and integration paths.}
\end{table}

The variation across ecosystems (Table~\ref{tab:wearable-import-rates}) reflects differences in the extent to which their proprietary data sources support importing data into the common ingestion layer. This is an important system-level consideration because the availability of wearable data via Google Health Connect underpins the downstream health information and interaction features across the application suite at the time of this testing.

\subsection{Descriptive Usability Results}

\subsubsection{System Usability Scale}

All 26 participants (13 subjects and 13 care actors) completed the ten-item System Usability Scale. The overall mean SUS score was $80.77$ ($SD=10.60$), with a median of $81.25$ and scores ranging from $62.5$ to $97.5$. Subject responses had a mean score of $81.92$ ($SD=11.23$), while care-actor responses had a mean score of $79.62$ ($SD=10.25$). Table~\ref{tab:sus-results} presents the descriptive statistics for the two participant roles and the combined sample.

\begin{table}[ht!]
\centering
\small
\begin{tabular}{lrrrrr}
\toprule
\textbf{Participant group} & \textbf{$n$} & \textbf{Mean} & \textbf{SD} & \textbf{Median} & \textbf{Range} \\
\midrule
Subjects & 13 & 81.92 & 11.23 & 82.50 & 65.0\textendash97.5 \\
Care actors & 13 & 79.62 & 10.25 & 77.50 & 62.5\textendash95.0 \\
\midrule
All participants & 26 & 80.77 & 10.60 & 81.25 & 62.5\textendash97.5 \\
\bottomrule
\end{tabular}
\caption{\label{tab:sus-results} Descriptive SUS scores for subjects, care actors, and the combined laboratory-study sample.}
\end{table}

\subsubsection{Component Usability Matrix}

The Component Usability Matrix captured participants' ratings of four interface components across visual clarity, ease of interaction, and perceived trustworthiness of information. All 26 participants completed the ratings. Mean scores for visual clarity ranged from $4.35$ to $4.50$, ease-of-interaction scores ranged from $4.31$ to $4.42$, and perceived information-trustworthiness scores ranged from $4.23$ to $4.27$ on the five-point scale. Table~\ref{tab:component-results} reports the mean and standard deviation for each component and rating dimension.

\begin{table*}[ht!]
\centering
\small
\resizebox{\textwidth}{!}{
\begin{tabular}{lccc}
\toprule
\textbf{Interface component} &
\textbf{Visual clarity} &
\textbf{Ease of interaction} &
\textbf{Information trustworthiness} \\
\midrule
Wearable telemetry dashboard & 4.50 (0.65) & 4.35 (0.63) & 4.23 (0.82) \\
Role-based sharing dashboard & 4.42 (0.70) & 4.42 (0.70) & 4.27 (0.83) \\
AI Verification Engine and confidence badges & 4.38 (0.64) & 4.35 (0.75) & 4.23 (0.76) \\
Access logs and revocation controls & 4.35 (0.75) & 4.31 (0.74) & 4.27 (0.72) \\
\bottomrule
\end{tabular}}
\caption{\label{tab:component-results} Component Usability Matrix ratings across the four evaluated interface components ($n=26$). Values are reported as mean (standard deviation) on a five-point Likert scale.}
\end{table*}

The wearable telemetry dashboard received the highest mean visual-clarity rating ($M=4.50$, $SD=0.65$). The role-based sharing dashboard received mean ratings of $4.42$ ($SD=0.70$) for both visual clarity and ease of interaction. The AI Verification Engine received a mean visual-clarity rating of $4.38$ ($SD=0.64$) and an ease-of-interaction rating of $4.35$ ($SD=0.75$). Ratings for perceived information trustworthiness ranged from $4.23$ for the wearable dashboard and AI Verification Engine to $4.27$ for the role-based sharing dashboard, access logs, and revocation controls.

\subsection{Qualitative Findings}

The thematic analysis produced five overarching themes and 23 second-order subthemes. Together, these themes characterize how participants understood collective care roles, interpreted health information and AI-generated responses, negotiated the social implications of sharing, and experienced the system's interaction and configuration features. The findings below retain the distinction between subject and care-actor accounts, including differences in how participants, both within and across dyads, interpreted the same features. Rather than treating these accounts as uniform, the analysis describes where participants' experiences converged and where they reflected different expectations, relationships, and information needs.

\subsubsection{Collective Care Roles and Shared Health Management}

Participants understood the Care Circle as a support network whose membership could extend beyond family relationships. When interpreting the Care Circle interface, \texttt{JLSPWB6S} noted that its membership could include friends as well as relatives:

\begin{quote}
    ``Within your circle, maybe not family, but maybe your friends.'' \textemdash\texttt{JLSPWB6S}
\end{quote}

This understanding of membership was accompanied by participants' attention to the visibility associated with each care role. Care actors described the shared dashboard in relation to the subject's ability to select what was disclosed and to view only the information that had been approved. As \texttt{JLSPWA1C} explained:

\begin{quote}
    ``She had the option to select what data she wanted to share, and then I could see the data that she already had approved to be shared.'' \textemdash\texttt{JLSPWA1C}
\end{quote}

Participants, however, differed in what they considered an adequate level of information for shared health management. Some care actors considered the available metrics sufficient for remaining informed. \texttt{JLSPWA2C}, for example, described the combination of sleep, activity, steps, heart rate, and SpO$_2$ as sufficient to understand another person's health. Others wanted more interpretive context alongside the measurements. When reviewing the dashboard, \texttt{JLSPWB6C} asked:

\begin{quote}
    ``For heart rate, if it's 80 bpm, is it good, is it bad? How much am I off from that normal range?'' \textemdash\texttt{JLSPWB6C}
\end{quote}

The difference was not necessarily about the number of metrics displayed, but about whether the information could be interpreted without additional explanation. The shared dashboard was also discussed as a means of remote oversight. One care actor described being able to inspect the information when needed rather than repeatedly asking the subject about their daily routines:

\begin{quote}
    ``I don't need to ask him what your sleep is, how much you have walked, or what activities you have done. I can check it from my end randomly whenever I need to check it.'' \textemdash\texttt{JLSPWB2C}
\end{quote}

Subjects, in turn, described disclosure as a decision that could be made at the level of individual metrics. \texttt{JLSPWA1S} explained that sharing did not require exposing the entire health profile:

\begin{quote}
    ``I don't have to share the entirety of the data, I can choose which data to share.'' \textemdash\texttt{JLSPWA1S}
\end{quote}

During the walkthrough, subjects selected and deselected individual metrics and checked whether those changes were reflected in the care actor's view. Participants also recognized that permissions could be time-bounded. \texttt{JLSPWA4S}, for example, identified the available sharing periods while configuring access:

\begin{quote}
    ``I want to share these metrics for a period of one month. I can change it to one week, six months, or a year.'' \textemdash\texttt{JLSPWA4S}
\end{quote}

These accounts show how participants distinguished the choice of what to disclose from the period for which that information would remain available. Metric selection and permission duration were therefore understood as separate aspects of managing access.

Participants' descriptions of care roles also reflected the responsibilities and decision-making arrangements within their relationships. One parent described monitoring as part of an ongoing parental responsibility:

\begin{quote}
    ``We take it as our duty as a parent to take care of him in every way.'' \textemdash\texttt{JLSPWA5C}
\end{quote}

Other subjects described health decisions as shared with particular care actors rather than delegated to them. \texttt{JLSPWB4S}, for example, explained her relationship with her husband in terms of shared decision-making:

\begin{quote}
    ``We basically decide together, so he is a part of the decisions.'' \textemdash\texttt{JLSPWB4S}
\end{quote}

Care actors also described the value of receiving information about the subject's health as a source of reassurance. \texttt{JLSPWA7C} explained that regular information about the subject's condition could reduce the need for repeated reminders and worry:

\begin{quote}
    ``I don't need to remind her again and again if she is doing... something is monitored on her app.'' \textemdash\texttt{JLSPWA7C}
\end{quote}

Thus, while subjects emphasized their ability to govern disclosure, care actors described shared information in relation to their responsibilities and their interest in remaining informed. These accounts situated the same system features within different but related roles in collective health management.

\subsubsection{AI-Supported Sensemaking, Provenance, and Trust Calibration}

Participants' accounts of AI-supported sensemaking involved both the traceability of wearable measurements and the evidence accompanying AI-generated responses. They considered where the underlying health data originated, whether references could be inspected, and how confidence indicators should be interpreted. These elements were not always equally visible or influential to participants.

For wearable measurements, participants used source labels to identify the application or device from which a value had been collected. \texttt{JLSPWA4S}, for example, identified Samsung Health as the source of the displayed measurement:

\begin{quote}
    ``The app name from which it was collected was Samsung Health.'' \textemdash\texttt{JLSPWA4S}
\end{quote}

Although participants could understand the source information once they located it, some had difficulty finding the corresponding control. \texttt{JLSPWB2S} explained that the source label was clear once found, but that locating it required effort. \texttt{JLSPWB3S} similarly remarked on the size of the source controls. The distinction between source clarity and source-control visibility was therefore present in participants' accounts: the information itself was understandable, while accessing it was not always straightforward.

Participants also examined the references accompanying the Verification Engine's responses. Several described the ability to open a source and inspect the underlying material as important to their assessment of the response. \texttt{JLSPWB4S}, for example, described the value of being able to examine the evidence independently:

\begin{quote}
    ``It would be able to give me evidence, and I would really like to cross-check that as well.'' \textemdash\texttt{JLSPWB4S}
\end{quote}

Other participants opened the supplied links during the walkthrough. \texttt{JLSPWA4S}, for instance, inspected a Mayo Clinic reference and described it as a legitimate destination. These accounts show that participants did not discuss citations only as labels accompanying the response; they also considered whether the references could be opened and examined.

Confidence indicators were interpreted alongside this source information, but participants did not assign them a uniform role in deciding whether to accept a response. Some considered the numerical rating a useful signal for judging how cautiously a claim should be approached. After receiving a rating of 3 out of 10, \texttt{JLSPWB3S} described it as a cue to reconsider the submitted claim:

\begin{quote}
    ``After seeing a 3 out of 10... it's nudging me to say that I have to change my perspective.'' \textemdash\texttt{JLSPWB3S}
\end{quote}

Other participants placed greater emphasis on the supporting evidence than on the score itself. As noted above, \texttt{JLSPWB4S} valued the opportunity to cross-check the references, while \texttt{JLSPWA5S} stated that the sources were more influential than the confidence rating. Participants therefore considered the numerical indicator alongside the sources and the plausibility of the response, rather than treating it as the sole basis for trust.

The Verification Engine was also used to examine health claims and distinguish among the different components of a submitted statement. After entering the claim that 15,000 steps per day were mandatory, \texttt{JLSPWB2S} described the response as identifying which parts were supported and what required attention:

\begin{quote}
    ``It's giving me what is true, what is false, and what should I focus on.'' \textemdash\texttt{JLSPWB2S}
\end{quote}

Participants also discussed the importance of contextualizing health-related responses to the individual's profile. Rather than receiving advice that applied uniformly to everyone, they valued responses that accounted for personal information, goals, and constraints. \texttt{JLSPWA7S}, when asked about the form of guidance that would be useful, stated:

\begin{quote}
    ``Less generic, more according to my data.'' \textemdash\texttt{JLSPWA7S}
\end{quote}

A related account came from \texttt{JLSPWA4S}, who discussed how the response incorporated information about chronic lower-back pain and allergies to ibuprofen and diclofenac. The participant noted that the recommendations took these constraints into account, including caution regarding high-impact exercise and suggestions for lower-impact alternatives. These accounts reflected an interest in advice that connected health information to the subject's stated circumstances and goals, rather than presenting decontextualized recommendations.

The need for contextualization was also evident in care actors' interpretation of the wearable measurements themselves. While some considered the dashboard's collection of metrics sufficient to stay informed, others wanted reference ranges and explanations to help them understand the significance of each value. The request from \texttt{JLSPWB6C} to understand whether a heart rate of 80 bpm was good or bad illustrates this distinction between seeing a measurement and being able to interpret it. Participants thus discussed AI-supported sensemaking across several connected but distinct activities: tracing the origin of wearable data, inspecting references, interpreting confidence indicators, examining health claims, and relating explanations to personal goals and constraints.

\subsubsection{Relational Friction, Privacy, and Surveillance Governance}

Participants described access logs as providing visibility into which care actors had viewed particular metrics and when those interactions had occurred. For some subjects, this visibility supported their sense of control by making the activities of other care-circle members inspectable. \texttt{JLSPWA1S}, for example, explained that knowing who had access to the information and what they were viewing could be reassuring:

\begin{quote}
    ``The more data you have, the calmer I get.'' \textemdash\texttt{JLSPWA1S}
\end{quote}

Other participants described a more ambivalent response. \texttt{JLSPWA5S} explained that being able to see how frequently a care actor accessed particular information provided some transparency, but could also create a feeling of being watched:

\begin{quote}
    ``I can see how frequently and what exactly they're seeing, so I'm being watched.'' \textemdash\texttt{JLSPWA5S}
\end{quote}

The same participant described the experience as providing both a sense of privacy and a loss of privacy: knowing what the care actor could see made access more transparent, while the continued availability of the information remained uncomfortable. Other subjects did not interpret access-log visibility as surveillance, particularly when the selected care actors were already trusted people with whom they were comfortable sharing information. Access-log auditability, therefore, did not elicit a uniform response; participants related its value and discomfort to their own expectations about access and monitoring.

Participants also described social pressure that could remain even when they had control over individual sharing permissions. Some anticipated that care actors might notice when a particular metric was withheld, making the decision not to share visible in itself. \texttt{JLSPWA5S} described this possibility as a source of pressure:

\begin{quote}
    ``Now they know that I'm not sharing a particular piece of information, so that makes me feel like I am compelled to do that.'' \textemdash\texttt{JLSPWA5S}
\end{quote}

Another subject distinguished this feeling from surveillance and described the expectation to act on shared health information as a form of accountability:

\begin{quote}
    ``Not surveillance, but it then becomes sort of a moral commitment.'' \textemdash\texttt{JLSPWB2S}
\end{quote}

The participant explained that sharing health information with a care actor could create an expectation of responding to subsequent advice or requests. These accounts show that control over the information being shared did not necessarily remove the social expectations that could accompany disclosure.

Participants also considered whether an AI-generated summary could mediate communication between subjects and care actors. Some perceived the summary as a neutral presentation of recorded information rather than a personal reminder or criticism from another person. \texttt{JLSPWA2S}, for example, described the summary as:

\begin{quote}
    ``It's a neutral way, not nagging or anything.'' \textemdash\texttt{JLSPWA2S}
\end{quote}

This perception was not shared uniformly across relationships. Participants who already experienced reminders as expressions of concern did not necessarily regard them as unwanted pressure. \texttt{JLSPWA1S}, for instance, described their family as not being nagging, while other participants described reminders as a form of care. A care actor in another dyad similarly characterized reminding a friend about health routines as an expression of support rather than an obligation. The acceptability of monitoring and reminders was therefore discussed in relation to existing trust, closeness, and established patterns of interaction.

The anticipated effect of AI-generated summaries also depended on how frequently they would be sent and how they were framed. Participants considered that a summary could reduce repeated questions by making information available to care actors, but also noted that frequent updates could become intrusive or easy to ignore. \texttt{JLSPWB2S} contrasted weekly and daily updates:

\begin{quote}
    ``If it's once a week, then it would feel more like care... If it's a daily update, then it would feel like nagging.'' \textemdash\texttt{JLSPWB2S}
\end{quote}

The participant further noted that sharing a summary could create expectations to respond to the information received. Thus, participants' accounts of AI-mediated communication involved both the possibility of a more neutral presentation of health information and the conditions under which automated updates might create additional social pressure. These were anticipated effects discussed during the laboratory session, rather than changes observed through sustained use.

\subsubsection{Interaction Legibility and Technical Reliability}

Participants' descriptions of the interface reflected both ease of navigation and points at which particular controls or settings required further explanation. Some care actors found the interface straightforward to navigate and considered the information boundaries clear. \texttt{JLSPWA1C}, for example, described the overall interface as intuitive:

\begin{quote}
    ``There's nothing that's particularly overwhelming about the app. The UI is very intuitive.'' \textemdash\texttt{JLSPWA1C}
\end{quote}

Other participants were less familiar with specific configuration options and needed to inspect them more closely to understand their purpose. While exploring one of the settings, \texttt{JLSPWA5S} explained:

\begin{quote}
    ``Not too used to this one, so I'll check... what all it means exactly.'' \textemdash\texttt{JLSPWA5S}
\end{quote}

These accounts illustrate variation in the interface's legibility: participants could navigate the main screens, but some technical settings were less immediately understandable than the primary dashboard and sharing functions.

Participants also commented on the information provided before interacting with the system. \texttt{JLSPWA5S} explicitly valued being informed about data collection and sharing at the outset:

\begin{quote}
    ``I love that at first, [it] tells me how my data will be collected, and how it will be shared.'' \textemdash\texttt{JLSPWA5S}
\end{quote}

This response concerned the initial orientation to data use, before the participant configured individual sharing permissions or reviewed the shared dashboard.

Participants' preferences for the amount and placement of information on the dashboard also varied. Some valued having several health measurements and summaries consolidated in one place. \texttt{JLSPWA5C}, for example, considered a single-screen summary useful for reviewing progress against a selected goal and planning the next day's activities. Others wanted more salient information to be visible immediately on the overview screen. \texttt{JLSPWB6C} asked for a more prominent summary of relevant measurements and insights:

\begin{quote}
    ``You should get a sneak peek in the overview screen itself.'' \textemdash\texttt{JLSPWB6C}
\end{quote}

These comments reflect different preferences for information density and glanceability. While a consolidated dashboard was considered useful, some participants wanted the most relevant information to be more immediately visible, with additional detail available when they navigated further into a metric.

\subsubsection{Personal Agency and Reflective Engagement}

Participants described the personal-goal and notification settings as ways to configure which aspects of their health they wanted the application to prioritize and how they wanted to receive updates. Some related these settings to specific goals, such as improving fitness or sleep, while others interpreted the care-plan options in terms of the conditions and frequency under which notifications would be generated. When reviewing the goal settings, \texttt{JLSPWB2S} connected the configuration to the tailoring of the application:

\begin{quote}
    ``This would cater according to my goal, is it? Like, for example, improving fitness or getting better sleep.'' \textemdash\texttt{JLSPWB2S}
\end{quote}

Participants also considered whether notifications should be associated with positive changes, negative changes, or particular thresholds. These accounts concerned the choices available for configuring personal goals and notifications, rather than demonstrating how those settings affected behavior over an extended period.

The application activity screen was interpreted by some participants as a record that they could use to reflect on their own engagement and areas of attention. Participants described reviewing which metrics or functions they visited most frequently and relating those patterns to their own concerns. \texttt{JLSPWA4S}, for example, considered how repeated visits to the sleep screen could indicate concern about sleep:

\begin{quote}
    ``If I'm watching my sleep, that I clicked on that screen for more than three times or four times, so I should... think that I'm more concerned that I'm not getting enough amount of sleep.'' \textemdash\texttt{JLSPWA4S}
\end{quote}

The participant described the activity record as making an existing concern visible rather than introducing a new one. Another subject similarly interpreted frequent visits to the heart-rate screen as an indication of what they might be worried about. Participants therefore discussed usage analytics as more than a record of application activity: they also considered how these records could help them recognize which health topics had received their attention.
\section{Discussion}
\label{sec:discussion}

The central challenge in engineering collective health systems is not simply enabling multiple people to access the same health information. It concerns the interdependent activities through which people disclose, interpret, question, and act on information within existing care relationships. This requires interactive systems to account for both the technical conditions of information exchange and the social conditions under which such exchange becomes acceptable and useful. Drawing on the design and evaluation of \textit{JEEVHITAA}, we discuss five implications for the HCI and health-informatics communities: treating collective care as a first-class interaction and engineering concern; understanding privacy as relational governance rather than access control alone; moving from confidence-oriented AI interfaces towards evidence-oriented, collaborative sensemaking; recognizing automated health communication as a social intervention; and evaluating interactive health systems through complementary layers of technical, interactional, and social evidence. Together, these perspectives suggest that collective health technologies need to be evaluated not only by whether their components function, but also by how their interaction mechanisms shape the relationships and interpretive practices in which they are embedded.

\subsection{From Shared Health Information to Collective Care Architectures}

Empirical findings from our dyadic study underscore that extending personal health technologies to collective settings requires rethinking their underlying data and interaction abstractions. While early self-tracking models centered on individual reflection and behavior change~\cite{10.1145/1753326.1753409}, subsequent scholarship has established that health management is routinely distributed across families and social networks, where data collection, interpretation, and action are divided among multiple actors~\cite{10.1145/2998181.2998362,10.1145/3274396,10.2196/10830,10.1093/jamia/ocac049,10.1145/3579545}. Existing systems have begun operationalizing multi-actor support through multi-device household tracking and AI-mediated updates in formal care facilities~\cite{10.1145/3686998,10.1145/3772318.3791277,10.1145/3832000,10.1145/3757599}. However, our participants' accounts reveal that simply exposing telemetry to multiple viewers does not automatically produce coordinated care; rather, \textit{systems must be structured to encode who is viewing the data, why they need it, and what boundaries govern that visibility.}

A critical engineering distinction therefore lies between making health records shareable and designing architectures that support collective sensemaking. As our findings showed, the same physiological metric can be sufficient for one care actor while leaving another uncertain about whether a reading warrants concern, echoing broader insights on common information spaces and collaborative interpretation~\cite{10.2196/10830,10.1145/2998181.2998328}. Moreover, because informal caregiving roles and informational needs fluctuate as health circumstances evolve~\cite{10.1145/3686998,10.1145/3772318.3791609,10.1145/3613904.3642728,10.1145/3817013}, treating multi-user access as an ad-hoc permission layer atop a single-user schema creates brittle interfaces and forces users into out-of-band workarounds. Collective health architectures must instead treat relationships, role-specific views, and interpretive scaffolding as coupled design primitives.

\textit{JEEVHITAA} demonstrates one architectural realization of this principle by binding a canonical subject profile to care-circle memberships, role-conditioned views, metric-level disclosure, and time-bounded permissions. Rather than introducing isolated privacy features, the architecture enforces consistency between backend access policies and frontend visibility cues so that information boundaries remain legible to both subjects and care actors. Unlike facility-oriented tools that synthesize updates across clinical staff and relatives~\cite{10.1145/3757599}, this approach places governance directly with the subject, allowing everyday wearable telemetry to be selectively partitioned across heterogeneous personal relationships.

At the same time, our study exposes an architectural boundary between \textit{information visibility} and \textit{care responsibility}. Granting a care actor read access to a metric does not specify whether they are expected to interpret anomalies, initiate follow-up conversations, or intervene—a tension reflected in prior work on co-responsibility and invisible caregiving labor~\cite{10.1145/3357236.3395469,10.1145/3772318.3791609}. Next-generation collective care architectures should therefore model responsibility states, actor availability, and escalation expectations alongside access permissions, particularly in multi-member care circles, where ambiguity about who will act is heightened beyond dyadic pairs.

Ultimately, this work argues for \textit{shifting the primary unit of health-system engineering from the individual account with optional sharing to the care relationship and its evolving information flows}. Supporting socially distributed care does not require imposing a uniform family dashboard on all users; rather, it calls for extensible architectures capable of representing diverse relational roles, asymmetrical information needs, and dynamic disclosure boundaries as first-class system constructs.

\subsection{Privacy Engineering Must Account for the Social Cost of Disclosure}

Privacy in collaborative health systems is often operationalized through authentication, authorization, and access-control policies. These mechanisms are essential, but they do not fully capture how people experience the disclosure of health information within relationships. Contextual integrity frames privacy as the appropriateness of information flows in relation to actors, information attributes, and transmission principles, rather than simply the presence or absence of access~\cite{Nissenbaum_2004}. Research on the non-static nature of patient consent similarly demonstrates that privacy expectations can change with health circumstances, technology use, and relationships with information recipients~\cite{10.1145/2441776.2441838}. Research on home data sharing has shown that preferred access policies can be multidimensional and difficult to specify through conventional controls, while mobile-sensing work has explored data inspection and context-sensitive mechanisms for adjusting consent~\cite{10.1145/1753326.1753421,10.1145/3613904.3642815}. In collaborative settings, privacy is therefore better understood as an ongoing governance process than as a one-time permission decision.

This distinction has a practical consequence for system design. Granular permissions and revocation can establish what the system allows, but they cannot determine whether a person feels comfortable refusing a request, withholding a metric, or changing a previously granted permission. In a care relationship, the absence of disclosure may itself communicate something. A subject who withholds a particular measurement may be perceived as concealing information, even when the system explicitly affords that choice. Consequently, the ability to control access does not necessarily translate into the social freedom to exercise that control. Work on caring through data has similarly shown that health information sharing can become intertwined with emotional expectations, accountability, and the experience of caregiving~\cite{10.1145/2998181.2998303}. The unintended consequences identified in personal informatics research further demonstrate that tracking and data visibility can impose emotional, cognitive, and social costs alongside their intended benefits~\cite{10.1145/3715336.3735746}.

This exposes a limitation in evaluating privacy mechanisms by focusing solely on policy compliance. A system may correctly enforce a subject's permissions while still creating conditions that compel subjects to share. Conversely, visibility into access histories may provide accountability to one participant while making another feel continuously observed. The concept of social translucence helps explain this tension: visibility and accountability can support coordinated social activity, but their consequences depend on the social practices and expectations that develop around them~\cite{10.1145/344949.345004}. Research on collaborative privacy practices has likewise shown that formal privacy policies do not always align with the practical realities of collaborative work, creating tensions around accountability and informal workarounds~\cite{10.1145/2531602.2531643}. In health settings, where information can be closely connected to personal identity and vulnerability, these tensions warrant particular attention.

For the HCI community, this suggests broadening the evaluation of privacy-preserving health systems beyond whether unauthorized access is prevented. Evaluation should also consider whether users understand the implications of granting access, whether declining or withdrawing access is socially feasible, and whether transparency mechanisms create unintended expectations of reciprocal visibility. The design implications extend beyond adding more permission settings. Systems should communicate the consequences of disclosure in understandable terms, distinguish between permission status and expectations of participation, and avoid interface cues that frame non-disclosure as an exceptional or undesirable state. Research on usable home access control and contextual privacy mechanisms reinforces the importance of making data flows and the consequences of control inspectable, rather than exposing policy settings alone~\cite{10.1145/1753326.1753421,10.1145/3613904.3642815}. Access histories should support accountability without making continuous monitoring the implicit norm.

An important research direction is to \textit{investigate the relationship between technical agency and experienced agency.} This requires examining not only whether users can configure or revoke permissions, but also whether they feel able to exercise those controls across different relationships and circumstances. Future studies should compare privacy mechanisms across care relationships, investigate how disclosure expectations change over time, and examine whether access transparency has different effects depending on the relative authority or dependency of the actors involved. Such work would help move privacy engineering from a predominantly rule-enforcement perspective towards one that accounts for the social conditions under which information governance is exercised.

\subsection{Towards Evidence-Oriented and Collaborative AI Sensemaking}

The integration of generative AI into health applications introduces a further challenge: the system is no longer simply presenting measurements but participating in their interpretation. Recent work has explored how generative AI can support personal health informatics activities such as planning, tracking, reflection, and action, while also identifying the need to account for users' goals, expertise, and contextual requirements~\cite{10.1145/3749503}. Systems such as HealthGuru~\cite{10.1145/3706598.3713852} and GLOSS~\cite{10.1145/3749474} extend this direction by using wearable or passive sensing data to support personalized health guidance and open-ended sensemaking across heterogeneous sources. Research on AI-assisted personal health decisions similarly characterizes trust as selective and dependent on the fit between AI recommendations and individual decision-making preferences~\cite{10.1145/3706598.3713462}. In a collective setting, however, interpretation may involve several people with different levels of familiarity with the subject's health, different responsibilities, and different requirements for evidence. This makes the design of AI-supported health interfaces a collaborative sensemaking problem, not solely a question of generating understandable responses for an individual user.

One implication is that the \textit{notion of AI transparency needs to be more carefully decomposed.} At least three forms of information are relevant in a health-information interface: \textit{the provenance of the underlying measurement, the evidence supporting an AI-generated interpretation, and the uncertainty associated with the system's assessment.} These are related but not interchangeable. A wearable data source label indicates where a measurement originated; a citation or evidence link enables examination of the basis for a claim; and a confidence indicator communicates an aspect of the system's assessment. Work on the lived experience of self-tracking metrics further shows that people interpret measures in relation to their own purposes and changing circumstances, rather than treating metrics as self-explanatory indicators~\cite{10.1145/3706598.3713650}. Research on consumer health information also distinguishes the related but non-identical concepts of information quality, credibility, and trust, reinforcing the need to specify what users are being asked to evaluate~\cite{10.1145/3576840.3578331}. Treating provenance, supporting evidence, and uncertainty as a single notion of ``trustworthiness'' risks obscuring what users are actually being asked to trust.

Human\textendash AI interaction research has emphasized that explanations should help users understand system behavior, assess its outputs, and determine when to rely on or question its recommendations~\cite{10.1145/3290605.3300233}. However, explanation does not necessarily produce appropriate reliance. Research on AI-assisted credibility assessment demonstrates that different forms of evidence and explanation can influence trust and reliance in different ways, while also exposing the possibility of automation bias and backfire effects~\cite{10.1145/3686922}. Work on user characteristics in explainable AI further cautions against assuming that personalization based on increasingly detailed user attributes will necessarily improve understanding~\cite{10.1145/3613904.3642352}. Together, these studies suggest that the \textit{central design objective should not be to just maximize trust in AI, but to support users in evaluating when an output is sufficiently grounded for the decision at hand.} Related health information systems, such as Evidotes, also illustrate how scientific evidence can be presented alongside experiential accounts to help users reason about health-related uncertainty~\cite{10.1145/3772318.3791089}.

For collective health systems, this also raises the question of whose understanding an explanation is intended to support. A subject may need to understand how a measurement was collected and whether it should be shared, while a care actor may need additional context to interpret the same measurement. A clinician, where involved, may require a different level of detail and evidence. These differences do not necessarily call for separate AI decisions; they may instead require different representations of the same underlying information, with clear boundaries around what each view communicates. Work on interactive model interpretation and domain-aligned medical explanations illustrates the value of making evidence and reasoning inspectable in forms that are meaningful to the intended user~\cite{10.1145/3544549.3585604,10.1145/3706598.3714058}. Research on hybrid explanations for lay users of mobile health recommendations further emphasizes the need to align explanation formats with users' comprehension and information needs~\cite{10.1145/3772071}. Recent systems for open-ended sensemaking over passive health data also demonstrate growing interest in supporting interpretation across heterogeneous information sources, although their evaluation does not remove the need to examine how people understand and use the resulting insights~\cite{10.1145/3749474}.

The implication is a \textit{move from confidence-led interfaces towards evidence-oriented interaction.} Rather than presenting a single confidence value as a proxy for reliability, systems should enable tracing generated claims to their underlying measurements or external sources, inspecting the evidence used in an interpretation, and understanding the limits of the system's assessment. Where personalized explanations are provided, they should make clear which user-specific information informs the explanation, rather than presenting personalization as an unexplained improvement in relevance. The aim is to support questioning and verification, not to encourage users to accept AI outputs because they appear authoritative.

This direction also establishes a more demanding evaluation agenda. Technical measures of source attribution, hallucination, and output conformance are necessary, but they do not establish whether users can identify unsupported claims, distinguish measurement provenance from evidence provenance, or appropriately revise their reliance when evidence is weak. Future studies should therefore combine independently annotated or expert-reviewed health-information cases with task-based evaluations of comprehension, evidence inspection, and decision-making. Comparative studies could examine whether users make better-informed judgments when source citations, confidence indicators, explanation formats, or combinations of these elements are included. In multi-actor settings, these assessments should also examine whether explanations support a shared understanding or lead to divergent interpretations among care-circle members.

\subsection{AI-Mediated Communication Is a Social Intervention}

Beyond interpreting information, AI-generated summaries and notifications can influence how health information enters interpersonal communication. This is particularly important because care actors and subjects do not interact with health information in a social vacuum. Personal informatics in interpersonal contexts has shown that data can mediate relationships and support or complicate existing social practices~\cite{10.1145/3274396}. Research on caring through data and collaborative caregiving also demonstrates that monitoring and information exchange are connected to emotional labor and the practical work of maintaining care relationships~\cite{10.1145/2998181.2998303,10.1145/3772318.3791609}. AI-enabled systems such as Family In The Loop further illustrate how generated updates can become part of communication between family members and care staff, rather than remaining isolated system outputs~\cite{10.1145/3757599}. More broadly, research on smart-home health systems has shown that introducing technology can change the distribution and nature of shared care work, sometimes creating additional work rather than simply reducing it~\cite{10.1145/3544548.3581546}.

These perspectives suggest that an \textit{automated summary should not be treated as a neutral container of information.} Its tone, timing, frequency, and intended audience can all affect how it is interpreted. A reminder delivered by a system may be perceived differently from the same reminder delivered directly by a family member; however, the system-generated message can still carry the expectations and authority associated with the relationship. Similarly, increasing the frequency of updates may improve information availability while also increasing the sense that a subject is expected to report or account for their behavior. \textit{The social consequences of automation therefore depend not only on the correctness of the generated content but also on the communication practices that the system enables.}

For system designers, this means notification delivery should be understood as part of interaction design and care governance, rather than as a downstream output of an automated pipeline. The relevant design choices include who receives a summary, whether the subject is aware of its distribution, how frequently updates are generated, whether the recipient can request additional information, and how the system communicates uncertainty or the absence of data. These decisions should remain distinguishable from the generation of the underlying health insight. A technically valid summary does not automatically imply that it should be sent, that it should be sent to every care actor, or that it should be delivered at the same frequency across relationships.

A related concern is that automation may relocate rather than eliminate coordination work. If a system produces more summaries, someone may still need to interpret them, decide whether they warrant a response, or explain them to another member of the care circle. Recent work on collaborative family caregiving highlights that tasks are not automatically shareable: delegation depends on the nature of the task and whether the effort of seeking or receiving help is justified~\cite{10.1145/3772318.3791609}. Consequently, future collective health systems should examine not only whether automated communication reduces repeated inquiries, but also who is expected to respond to system-generated information and whether the system creates new forms of monitoring, explanation, or follow-up work.

This is an important distinction for the HCI community. \textit{The opportunity is not to replace interpersonal communication with an ostensibly neutral AI intermediary, but to design AI-mediated communication that remains configurable, contestable, and sensitive to the relationships in which it operates.} Future evaluations should investigate how message cadence, tone, and audience affect perceived pressure, responsiveness, and the distribution of communication work. Such evaluations should include repeated interactions in everyday settings, since a notification acceptable during a controlled walkthrough may take on a different meaning when it becomes part of a recurring routine.

\subsection{Engineering Interactive Health Systems Requires Layered Evidence}

The preceding discussion points to a broader implication for engineering interactive health systems: evaluation needs to connect technical correctness with interactional and social consequences. The EICS community explicitly considers the design and engineering of interactive computing systems throughout their lifecycle, making the integration of implementation, verification, validation, and user-facing interaction central to its contributions~\cite {10.1145/3300960,10.1145/3457140}. In health applications, this integration is particularly consequential because data ingestion, access control, generated interpretation, and user interaction can each introduce different classes of failure.

The evaluation of \textit{JEEVHITAA} illustrates the value of treating these as distinct but connected evidence layers. Synthetic-profile tests can assess whether role-conditioned outputs conform to expected information scopes; access-control checks can verify that configured visibility boundaries are enforced; and integration tests can confirm that data can be imported across supported device ecosystems. These forms of system-level testing are important foundations for reliable interactive health applications. Prior mHealth engineering work has similarly emphasized secure wearable architectures and configurable mobile platforms for health data collection and processing~\cite{10.1145/2676431.2676432,10.1145/3300061.3345433}. However, conformance to a synthetic reference does not establish that a health interpretation is clinically appropriate, just as successful data import does not establish that the originating sensor measurement is accurate.

A complementary layer concerns the user's ability to understand and act on the system's behavior. Usability measures provide information about perceived ease of use, but collaborative health systems also require evidence about whether users understand information boundaries, interpret AI-supported explanations, and can exercise control over disclosure. The distinction is important because an interface can be perceived as usable while leaving uncertainty about the meaning or consequences of an action. This reinforces the need to combine conventional usability evaluation with task-based comprehension and evidence-inspection assessments, especially when systems mediate access to sensitive information or present AI-generated health interpretations.

A further layer, which remains open for future work, concerns\textit{ how these mechanisms operate in everyday care arrangements.} The wider HCI literature has increasingly documented how personal informatics and collaborative health technologies can produce both intended benefits and unintended consequences, including additional cognitive, emotional, and coordination work~\cite{10.1145/3715336.3735746,10.1145/3544548.3581546,Abdelaziz_Garfield_Neves_Lloyd_Norton_van_Dael_Wheeler_McLeod_Franklin_2024}. This means that successful engineering cannot be reduced to functional compliance or positive usability ratings. Field evaluations need to investigate whether information is useful to the people receiving it, whether sharing remains acceptable as circumstances change, and whether automated communication changes the amount or distribution of work involved in care.

For EICS and the broader health-HCI community, this suggests a layered evaluation strategy: (1) \textit{verify technical conformance and access-policy enforcement;} (2) \textit{validate the semantic and clinical quality of generated health information against independently reviewed references;} (3) \textit{assess users' understanding, evidence inspection, and ability to exercise control;} and (4) \textit{examine the longer-term relational and operational consequences of deploying the system in diverse care arrangements.} These layers are complementary rather than substitutable. A system that performs well at one layer may still fail at another, and the evidence required to make claims at each layer differs. The value of an integrated system evaluation lies in making these distinctions visible rather than collapsing them into a single measure of system effectiveness.

\subsection{Limitations and Future Research}

The scope of the present evaluation places important boundaries on the conclusions and identifies where stronger evidence is needed. The laboratory study involved 13 subject\textendash care-actor dyads, with subjects aged 18\textendash 37; it therefore does not establish the usability or suitability of \textit{JEEVHITAA} for older adults, despite the system's broader motivation around older-adult care. This gap matters because prior research with older adults has identified concerns about privacy, security, and the complexity of health-monitoring controls, while studies of informal care coordination highlight the uncertainty and distributed responsibilities that can shape support arrangements~\cite{10.1145/3706598.3714019,10.1145/3817013}. The dyadic study configuration also leaves open how the interaction mechanisms operate when several care actors have different roles, access permissions, and interpretations of the same information. Future studies should directly involve older adults and evaluate larger, more diverse care circles, including relationships beyond the immediate family.

The controlled walkthrough provides an initial evaluation of interaction mechanisms, but cannot establish how disclosure preferences, notification expectations, or interpretations of AI-generated information evolve through repeated use. Longitudinal field studies are needed to examine the stability of permissions, the social consequences of access transparency, and whether automated summaries alter communication and coordination work. Such studies should measure both intended outcomes and potential costs, rather than treating engagement or information availability as sufficient indicators of benefit. The dyadic interview format and single-primary-researcher thematic analysis also warrant consideration when interpreting the qualitative findings and transferring them to other care contexts.

Finally, the technical evaluation used synthetic care-circle profiles and programmatically constructed references. This supports repeatable assessment of role-specific information and access constraints, but does not provide clinically validated ground truth. Stronger evaluation will require independently annotated health cases, expert assessment where appropriate, and task-based measures of users' ability to inspect evidence and recognize uncertainty. Wearable integration should likewise be evaluated separately for data availability, measurement quality, and the reliability of downstream interpretation. These distinctions will be essential before systems of this kind can be evaluated for clinical utility or deployed in settings where people may make consequential health decisions.

Taken together, the contribution of \textit{JEEVHITAA} is to demonstrate how the engineering of collective health interactions can bring role-conditioned access, privacy controls, provenance, and AI-supported interpretation into a single interactive system. More broadly, the findings motivate a research agenda in which collective health technologies are assessed not only as information systems, but as systems that participate in the governance of disclosure, the interpretation of evidence, and the organization of interpersonal communication. Moving this agenda forward requires stronger technical and clinical validation, more diverse multi-actor evaluations, and sustained investigation of how these systems fit within the relationships and routines they are intended to support.
\section{Conclusion}

We presented \textit{JEEVHITAA}, an implemented HCAI system designed to support collective health management by treating the care circle as a unit of interaction and information governance. Its contributions include an integrated approach to role-conditioned health information sharing, fine-grained and time-bounded access control, auditable information flows, wearable-data integration, and AI-supported health summaries and verification. Through technical evaluations across 1,000 synthetic care circles and a controlled laboratory study with 13 subject\textendash care-actor dyads, we examined the system's technical conformance and user-facing interactions. The study highlights the importance of designing not only for information access, but also for the privacy, interpretation, and communication needs that arise within care relationships. Future work should explicitly investigate accessibility for older adults and the longer-term implications of these mechanisms in diverse, real-world care circles.

\bibliographystyle{ACM-Reference-Format}
\bibliography{references}
\clearpage 





\appendix

\section{LLM Algorithms}
\label{appendix:algo}
The detailed algorithms for the verifier and summarizer are presented below.

\begin{algorithm}[ht!]
\caption{AAlgorithm for Profile Summary Generation for Subject \& Care Circle Actors}
\label{alg:profile_summary}
\begin{algorithmic}[1]
  \Require canonical profile $\mathcal{P}$ for subject $s$, care circle $G_s$, templates $T$, time window $W$
  \Ensure per-actor summaries $\{S_a\}_{a\in G_s}$

  \State $Features \gets \text{EnrichFeatures}(\mathcal{P},W)$
  \State $S_s \gets \text{TemplateFill}(T[\text{subject}],Features)$
  \State $S_s.conf \gets \text{EstimateConfidence}(Features)$
  \State \text{StoreSnapshot}(s,$S_s$); \ \text{IndexEmbedding}(s,$S_s$)
  \ForAll{actor $a$ \textbf{in} $G_s$}
    \State $role \gets G_s.role(a)$
    \State $Perm \gets \text{ResolvePermissions}(a,\mathcal{P},G_s)$
    \State $Fields \gets \text{FilterFields}(Features,Perm)$
    \State $brief \gets \text{TemplateFill}(T[role],Fields)$
    \State $brief.actionable \gets \text{ExtractTasks}(brief)$
    \State \text{AnnotateProvenance}(brief)
    \State \text{StoreSnapshot}(a,brief); \ \text{IndexEmbedding}(a,brief)
    \State \text{SendNotificationIfNeeded}(a,brief)
  \EndFor
  \State \Return $\{S_a\}_{a\in G_s}$
\end{algorithmic}
\end{algorithm}

\begin{algorithm}[ht!]
\caption{Algorithm for Health Information Verification Engine with User's Contextual Relevance}
\label{alg:context_builder}
\begin{algorithmic}[1]
  \Require user query $Q$, user id $u$, chat context $C$, profile store $\mathcal{P}$, care-graph $G$, retrieval size $k$
  \Ensure structured response context $CTX$ (snapshot, evidence IDs, prompts)

  \State $A \gets \text{ActorProfile}(u,G)$
  \State $Allowed \gets \text{ApplyACLs}(A,Q)$
  \State $Snapshot \gets \text{BuildSnapshot}(\mathcal{P},Allowed,C)$
  \Statex \hspace*{6em}$\triangleright$ \textit{privacy-limited}
  \State $E \gets \text{RetrieveTopK}(\text{Embed}(Snapshot),k)$
  \State $Sys \gets \text{SystemPersona}(A.role)$
  \State $User \gets \text{Concat}(Q,\text{ShortContext}(Snapshot),E)$
  \State $Draft \gets \text{LLM\_Generate}(\text{FormatPrompt}(Sys,User))$
  \State $Claims \gets \text{ExtractClaims}(Draft)$
  \ForAll{claim $c$ \textbf{in} $Claims$}
    \State $V \gets \text{LLM\_Verify}(c,\text{FindEvidence}(c,E,Snapshot))$
    \State $\text{AttachVerification}(Draft,c,V)$
  \EndFor
  \State $CTX \gets \text{SynthesizeContext}(Draft,E,Snapshot)$
  \State \Return $CTX$
\end{algorithmic}
\end{algorithm}
\clearpage

\section{Thematic Analysis Codebook}
\label{appendix:coding}

\footnotesize
\begin{longtable}{@{}p{0.13\textwidth} p{0.16\textwidth} p{0.21\textwidth} p{0.30\textwidth} p{0.14\textwidth}@{}}
\caption{\label{tab:codebook} Thematic analysis codebook documenting the five overarching themes and 23 second-order subthemes. Subthemes are marked as deductive [D] or inductive [I] according to their analytical origins. Themes 4 and 5 are identified as inductive themes. The table presents code definitions, representative participant evidence, and participant identifiers.}\\

    \toprule
    \textbf{Overarching Theme} & \textbf{Subtheme/Code} & \textbf{Code Definition} & \textbf{Representative Evidence} & \textbf{Participant IDs} \\
    \midrule
    \endfirsthead

    \multicolumn{5}{@{}l}{\textit{Table \ref{tab:codebook} (continued)}} \\
    \toprule
    \textbf{Overarching Theme} & \textbf{Subtheme/Code} & \textbf{Code Definition} & \textbf{Representative Evidence} & \textbf{Participant IDs} \\
    \midrule
    \endhead

    \midrule
    \multicolumn{5}{r}{\textit{(continued on next page)}} \\
    \endfoot

    \bottomrule
    \endlastfoot

    \textbf{Collective Care Roles \& Shared Health Management} & \textbf{Care Circle Membership and Role Legibility [D]} & How participants understood the Care Circle as a chosen support network and interpreted what an assigned care actor could see. This directly addresses the shift from individual management to multi-actor care and the understanding of roles/visibility. & \textit{JLSPWB6S: “within your circle, maybe not family, but maybe your friends”} \newline \textit{JLSPWA1C: “she had the option to select what data she wanted to share, and then I could see the data that she already had approved to be shared.”} & JLSPWB6S; JLSPWA1C; JLSPWA2S; JLSPWB7S \\ \addlinespace
     & \textbf{Shared Dashboard Adequacy and Remote Oversight [D]} & Whether the shared dashboard presents an appropriate and sufficient level of detail for the care actor, and whether remote access supports understanding without requiring continuous expert help. & \textit{JLSPWA2C: “sleep cycle, activity, steps, heart rate, SpO2 … it’s more than sufficient to know about a person’s health”; contrast with JLSPWB6C asking what an 80 bpm heart rate means.} \newline \textit{JLSPWB2C: “I don’t need to ask him what is your sleep, how much you have walked … I can check it from my end randomly whenever I need to check it.”} & JLSPWA2C; JLSPWB2C; JLSPWB6C; JLSPWA5C \\ \addlinespace
     & \textbf{Subject-Led, Metric-Level Disclosure [D]} & The subject configures which individual metrics are visible and adapts disclosure to particular recipients or situations rather than treating sharing as all-or-nothing. & \textit{JLSPWA4S: “I’ll now remove sleep”; “it’s removed from my friend’s screen as well.”} \newline \textit{JLSPWA1S: “if I don’t want my parents to know this bad for a week before exam, so I can remove it.”} & JLSPWA1S; JLSPWA4S; JLSPWA5S; JLSPWB4S; JLSPWB7S \\ \addlinespace
     & \textbf{Time-Bounded Permissions and Revocability [D]} & Participants recognized access duration and the ability to modify or revoke permissions as mechanisms for governing the temporal scope of sharing. & \textit{JLSPWA4S: “I want to share these metrics for a period of 1 month. I can change it to 1 week, 6 months, or a year.”} & JLSPWA4S; JLSPWA6S; JLSPWB2S; JLSPWB7S \\ \addlinespace
     & \textbf{Care-Role Responsibility and Remote Reassurance [I]} & Beyond interface comprehension, care actors described monitoring as a responsibility or a means of remaining reassured, while subjects differentiated the responsibilities of partners, parents, and friends. & \textit{JLSPWA4S: “we basically decide together, so he is a part of the decisions.”} \newline \textit{JLSPWA5C: “we take it as our duty as a parent to take care of him in every way”; JLSPWA7C said shared information would reduce worry.} & JLSPWA4S; JLSPWA5C; JLSPWA7C; JLSPWB2C; JLSPWB5C \\ \addlinespace
    \textbf{AI-Supported Sensemaking, Provenance \& Trust Calibration} & \textbf{Wearable Data-Origin Traceability [D]} & Participants used provenance labels to identify the app or device from which a health metric originated; some found the label or control difficult to locate. & \textit{JLSPWA4S: “the app name from which it was collected was Samsung Health”; JLSPWB2S: “I know what is the source of information.”} \newline \textit{JLSPWB2S: “Finding it was difficult where the source button is, but it is clear”; JLSPWB3S: “it’s hard, it’s tiny source buttons”.} & JLSPWA1S; JLSPWA4S; JLSPWB2S; JLSPWB3S; JLSPWA5C \\ \addlinespace
     & \textbf{Citation Inspectability and Cross-Source Verification [D]} & Trust was strengthened when participants could open cited links and verify source legitimacy; several still preferred independent cross-checking rather than treating AI output as self-validating. & \textit{JLSPWA4S: “the link it is providing is actually a legit link, and it is not leading to a fake link.”} \newline \textit{JLSPWA5S: “sources worked more for me than the confidence rating”; JLSPWB1S: “can’t just rely on one instance.”} & JLSPWA1S; JLSPWA4S; JLSPWA5S; JLSPWB1S; JLSPWB2S \\ \addlinespace
     & \textbf{Confidence-Score Interpretation and Trust Calibration [D]} & Participants treated scores as useful signals but did not uniformly defer to them: score meaning was weighed against cited evidence, prior knowledge, and whether the response was plausible. & \textit{JLSPWB3S: described a “3 out of 10” rating as a quantifier that showed whether the belief or question should be followed or avoided.} \newline \textit{JLSPWA4S: “If the personalized advice was in a different direction … the trust would have gone much less”; JLSPWA5S said sources worked more than the confidence rating.} & JLSPWA4S; JLSPWA5S; JLSPWB2S; JLSPWB3S; JLSPWB7S \\ \addlinespace
     & \textbf{AI Verification of Circulating Health Claims [D]} & Participants used the verification interaction to assess claims and distinguish supported, unsupported, or misleading advice, including claims encountered through online or messaging channels. & \textit{JLSPWB2S: “it is giving me what is true, what is false, and what should I focus on.”} & JLSPWA1S; JLSPWA2S; JLSPWA4S; JLSPWB2S; JLSPWB3S; JLSPWB7S \\ \addlinespace
     & \textbf{Personalized and Context-Sensitive AI Reasoning [I]} & Participants valued advice that incorporated age, profile, health conditions, allergies, goals, and daily context, and resisted generic or decontextualized advice. & \textit{JLSPWA4S: “it is personalizing the advice based on my experience” and accounts for the stated allergy to ibuprofen and diclofenac.} \newline \textit{JLSPWA1S: “it does not make serious something that is not serious”; JLSPWB7C: “it should not be generic. It should be more objective.”} & JLSPWA1S; JLSPWA4S; JLSPWA5S; JLSPWB3S; JLSPWB7C \\ \addlinespace
     & \textbf{Metric Comprehension and Health-Literacy Scaffolding [I]} & Care actors could identify displayed measurements but asked for reference ranges, interpretation, and practical context to understand whether values were meaningful or concerning. & \textit{JLSPWB6C: “For heart rate, if it’s 80 bpm, is it good, is it bad? How much am I off from that normal range?”} & JLSPWA2C; JLSPWB2C; JLSPWB6C; JLSPWB7C \\ \addlinespace
     & \textbf{Actionable AI Guidance and Goal Alignment [I]} & Participants valued recommendations when they translated a summary into feasible next steps that aligned with their own goals and constraints. & \textit{JLSPWA4S described suggestions to increase activity while accounting for lower-back pain and to choose low-impact activities.} & JLSPWA1S; JLSPWA4S; JLSPWB3S; JLSPWB7S \\ \addlinespace
    \midrule
    \textbf{Relational Friction, Privacy \& Surveillance Governance} & \textbf{Access-Log Auditability and Transparency [D]} & The access log made it possible to inspect who viewed particular data and when; some participants interpreted this visibility as reinforcing privacy and control. & \textit{JLSPWB3S: “at 4:53 he checked my heart rate, activity, oxygen saturation”; JLSPWA4S referred to how many times a care actor viewed each metric.} \newline \textit{JLSPWB3S: “Maintains my sense of privacy, as if I can see what … accesses my records”; JLSPWA1S said knowing who had access made them calmer.} & JLSPWA1S; JLSPWA4S; JLSPWB3S; JLSPWB7S \\ \addlinespace
     & \textbf{Surveillance Ambivalence and Self-Conscious Disclosure [D]} & Participants described opposing responses to visibility: the log could clarify access, but repeated access or visibility into viewing patterns could also make subjects or care actors feel watched or self-conscious. & \textit{JLSPWA5S: “I can see how frequently and what exactly they’re seeing, so I’m being watched”; JLSPWA4S raised concern about heart-rate monitoring every 15 minutes.} & JLSPWA4S; JLSPWA5S; JLSPWA4C \\ \addlinespace
     & \textbf{Disclosure Pressure and Accountability [D]} & Granular controls did not remove every social consequence: participants anticipated pressure to share withheld metrics or felt accountable to respond once health information had been shared. & \textit{JLSPWA5S: “now they know that I’m not sharing a particular piece of information, so that makes me feel like I am compelled to do that.”} \newline \textit{JLSPWB2S: “Not surveillance, but it then becomes sort of a moral commitment.”} & JLSPWA1S; JLSPWA5S; JLSPWB2S; JLSPWB4C \\ \addlinespace
     & \textbf{AI Neutrality as a Mediating Interaction [D]} & A neutral, descriptive AI briefing was sometimes seen as less interpersonal and less nagging than direct reminders; participants expected shared summaries to reduce repeated reminders in some relationships. & \textit{JLSPWB2S: “It just feels objective, like whatever it has recorded, it’s just giving it to me.”} \newline \textit{JLSPWA2S: “I think it would reduce the nagging, because they’ll be like actively [aware of] what’s going on”; JLSPWA7C expected fewer repeated reminders.} & JLSPWA1S; JLSPWA2S; JLSPWA7C; JLSPWB2S; JLSPWB3S \\ \addlinespace
     & \textbf{Relationship-Conditional Acceptance of Monitoring [I]} & The same access or monitoring feature was interpreted differently depending on trust, closeness, and established family or friendship norms; participants did not describe one universal privacy response. & \textit{JLSPWA1S: “My parents are not naggy”; JLSPWA6S: “there is no privacy concern” but anticipated social pressure from calls about eating.} & JLSPWA1S; JLSPWA4S; JLSPWA6S; JLSPWB3S; JLSPWB4S \\ \addlinespace
     & \textbf{Notification Cadence and Summary Tone as Friction Conditions [I]} & Frequency and framing shaped whether automated updates were perceived as caring or nagging; daily repetition could become ignorable or intrusive, and negative summaries could invite more questioning. & \textit{JLSPWB2S: “if it’s once a week … it would feel more like care”; “If it’s a daily update, then it would feel like a nagging.”} & JLSPWB1S; JLSPWB2S; JLSPWA4S; JLSPWA6S \\ \addlinespace
    \midrule
    \textbf{Interaction Legibility \& Technical Reliability} & \textbf{Navigation and Configuration Legibility [I]} & The core interface was often described as intuitive, but some configuration labels and technical concepts were unfamiliar or required explanation. & \textit{JLSPWA1C: “The UI is very intuitive”; JLSPWA4C said they could “easily navigate through the different screens.”} \newline \textit{JLSPWA5S: “Not too used to this one, so I’ll check … what all it means exactly”; JLSPWB6S asked what AI model selection meant.} & JLSPWA1C; JLSPWA4C; JLSPWA5S; JLSPWB1S; JLSPWB6S \\ \addlinespace
     & \textbf{Up-Front Data-Use Orientation [I]} & A participant explicitly valued being told before interacting with the system how data would be collected and shared, suggesting that privacy orientation is part of usability and acceptance. & \textit{JLSPWA5S: “I love that at first, [it] tells me how my data will be collected, and how it will be shared.”} & JLSPWA5S \\ \addlinespace
     & \textbf{Glanceable Dashboard and Information Density [I]} & Participants valued a clean, concise, and consolidated view, while also asking for salient insights on the landing screen and additional detail when needed. & \textit{JLSPWA5C: “it’s good to have all information in on a single dashboard”; JLSPWB6C wanted a “sneak peek” on the overview screen.} & JLSPWA5S; JLSPWA5C; JLSPWB2S; JLSPWB6C \\ \addlinespace
    \textbf{Personal Agency \& Reflective Engagement} & \textbf{Personal Goal and Notification Configuration [I]} & Participants valued the ability to set activity goals, choose reporting periods, and select when or for what kind of change they receive notifications. & \textit{JLSPWA1S selected which positive and negative changes should trigger notifications; JLSPWB2S interpreted care plans as controlling notification frequency.} & JLSPWA1S; JLSPWA2S; JLSPWB2S; JLSPWB7S \\ \addlinespace
     & \textbf{Usage Analytics as Self-Reflection [I]} & Interaction history was interpreted not only as system telemetry but as a mirror of the user's own attention, concerns, and health priorities. & \textit{JLSPWB2S: if heart rate is the most-visited screen, “it means I’m worried about it”; JLSPWA1S said the log showed what the care actor was checking repeatedly.} & JLSPWA1S; JLSPWA4S; JLSPWB2S; JLSPWB3S \\
\end{longtable}

\section{Usability Questionnaire}
\label{appendix:sus}

The following System Usability Scale (SUS) was administered to the Subject and Care Actors to evaluate the usability of the JEEVHITAA system. Participants rated each item on a 5-point Likert scale ranging from ``Strongly Disagree'' (1) to ``Strongly Agree'' (5).

\begin{framed}
\begin{enumerate}
    \item I think that I would like to use JEEVHITAA frequently.
    \item I found JEEVHITAA unnecessarily complex.
    \item I thought JEEVHITAA was easy to use.
    \item I think I would need support from a technical person to use JEEVHITAA.
    \item I found the various functions in JEEVHITAA (e.g., dashboards, AI verification) were well integrated.
    \item I thought there was too much inconsistency in this system.
    \item I would imagine that most people would learn to use JEEVHITAA very quickly.
    \item I found JEEVHITAA very cumbersome to use.
    \item I felt very confident using JEEVHITAA.
    \item I needed to learn a lot of things before I could get going with this system.
\end{enumerate}
\end{framed}

\section{Component Usability Matrix}
\label{appendix:componentusability}

The Component Usability Matrix was administered individually to Subjects and Care Actors following the interactive walkthrough. Participants evaluated four interface components across three dimensions: visual clarity, ease of interaction, and perceived trustworthiness of information. For each component, participants selected one rating for each dimension on a five-point scale, where 1 represented the least favorable rating and 5 the most favorable rating.

\begin{table*}[ht!]
    \centering
    \small
    \resizebox{\textwidth}{!}{
    \begin{tabular}{@{}p{0.31\textwidth}p{0.22\textwidth}p{0.22\textwidth}p{0.22\textwidth}@{}}
        \toprule
        \textbf{Interface component} & \textbf{Visual clarity} & \textbf{Ease of interaction} & \textbf{Information \newline trustworthiness} \\
        \midrule
        Wearable Telemetry Dashboard (24-hour Watch Ultra summary) & 1 \quad 2 \quad 3 \quad 4 \quad 5 & 1 \quad 2 \quad 3 \quad 4 \quad 5 & 1 \quad 2 \quad 3 \quad 4 \quad 5 \\
        \addlinespace
        Role-Based Sharing Dashboard (Access Control List / ACL) & 1 \quad 2 \quad 3 \quad 4 \quad 5 & 1 \quad 2 \quad 3 \quad 4 \quad 5 & 1 \quad 2 \quad 3 \quad 4 \quad 5 \\
        \addlinespace
        AI Verification Engine and Confidence Badges & 1 \quad 2 \quad 3 \quad 4 \quad 5 & 1 \quad 2 \quad 3 \quad 4 \quad 5 & 1 \quad 2 \quad 3 \quad 4 \quad 5 \\
        \addlinespace
        Audit Logs and Instant Revocation Action & 1 \quad 2 \quad 3 \quad 4 \quad 5 & 1 \quad 2 \quad 3 \quad 4 \quad 5 & 1 \quad 2 \quad 3 \quad 4 \quad 5 \\
        \bottomrule
    \end{tabular}}
    \caption{\label{tab:component-usability-instrument}Component Usability Matrix administered to participants.}
\end{table*}

For visual clarity, lower ratings indicate that the component was difficult to understand visually, and higher ratings indicate that it was visually clear. For ease of interaction, lower ratings indicate the component was difficult to use, and higher ratings indicate it was easy to use. For information trustworthiness, lower ratings indicate less perceived trustworthiness and higher ratings indicate greater perceived trustworthiness.

\section{Post-Walkthrough Semi-Structured Interview Guide}
\label{appendix:interviewguide}

The semi-structured interview was conducted with the Subject and Care Actor together after the interactive walkthrough. The guide covered participants' existing health-support practices and their evaluations of the system. Questions were directed to both participants unless otherwise indicated; role-specific prompts were used to elicit the distinct perspectives of the Subject and the Care Actor. Follow-up questions were used as needed to clarify responses and invite examples from the interaction. The questions listed in the guide are only sample probes and were not asked verbatim; the sequence depended on the flow of the interview.

\begin{framed}
\subsection{Part A: Existing Health-Support Practices}

\begin{enumerate}
    \item \textbf{Care circle and roles:} Think about the last time you needed help with a health-related task. Whom did you ask for help? What was their role, and how did you decide that they were the right person to involve?
    \item \textbf{Distribution of health-related activities:} Can you describe a typical week of health-related activities? Who usually checks health measurements, tracks medication or activity, and makes decisions about seeking further help?
    \item \textbf{Health information sources:} How and from whom do you usually receive health advice (for example, messaging applications, family members, or online sources)? How do you decide whether a health tip received on your phone is reliable or safe to follow?
    \item \textbf{Reminders and monitoring:}
    \begin{itemize}
        \item \textit{To the Subject:} Do reminders or questions from family members or friends about your health ever feel like nagging or monitoring? Could you describe an example?
        \item \textit{To the Care Actor:} How do you feel when you need to remind or ask the Subject about their health, medication, or activity routines?
    \end{itemize}
\end{enumerate}

\subsection{Part B: System Interaction and Experience Following the Walkthrough}

\subsubsection{I. Collective Care Roles and Information Sharing (RQ1)}

\begin{enumerate}
    \item \textbf{Role visibility:} Looking at the role-specific summary shown to [Care Actor Name], does the level of information shared match what you would want them to know? Is there anything you would add or prefer to keep private?
    \item \textbf{Private and shared information:} How easy was it to understand which health information was visible to the Care Actor and which information remained private to the Subject? Was this distinction clear in the interface?
    \item \textbf{Shared dashboard adequacy:} \textit{To the Care Actor:} Does the summary provide the level of detail you would need to understand the Subject's shared health information, or does it feel insufficient or overwhelming? What additional context, if any, would help?
    \item \textbf{Metric selection and access duration:} \textit{To the Subject:} How did you find selecting individual metrics and setting the duration of access? Are there circumstances in which you would change what is shared or how long it remains available?
\end{enumerate}

\subsubsection{II. AI-Supported Sensemaking and Provenance (RQ2)}

\begin{enumerate}
    \item \textbf{Wearable-data provenance:} When reviewing the 24-hour summary, how clear was it whether a particular measurement originated from the Samsung Galaxy Watch Ultra or from a manual entry? Was the source information easy to locate?
    \item \textbf{Verification and supporting evidence:} When you used the AI Verification Engine to assess the health tip, how did you interpret its result? What role did the confidence information and supporting citations play in your assessment?
    \item \textbf{Independent verification:} Did you inspect any of the cited sources? What did you look for when deciding whether a source was relevant or credible?
    \item \textbf{Contextual relevance:} How relevant was the generated explanation or guidance to the Subject's profile, health circumstances, or stated goals? What information would make the guidance more useful?
    \item \textbf{AI as an information intermediary:} How did the AI-generated briefing feel in comparison with receiving the same information directly from the Care Actor or another person? Did it come across as a neutral presentation of information, or did it create a different impression?
\end{enumerate}

\subsubsection{III. Privacy, Transparency, and Relational Experience (RQ3)}

\begin{enumerate}
    \item \textbf{Access history and transparency:} What did you think of being able to inspect the access history? Did seeing who accessed particular health information, and when, affect your sense of control or comfort?
    \item \textbf{Revocation and sharing control:} \textit{To the Subject:} How did knowing that you could change or immediately revoke access affect your willingness to share selected health information? Are there circumstances in which you would hesitate to use these controls?
    \item \textbf{Perceived monitoring:} Did any part of the system make you feel watched or monitored? Were there aspects of the access history or sharing controls that made you feel more in control? Please explain.
    \item \textbf{Reminders and communication:} How do you think system-generated health summaries or updates would affect the way you communicate about health? Could they reduce repeated questions, create additional expectations, or have a different effect?
    \item \textbf{Notification cadence and tone:} Would the frequency or wording of an update affect how you receive it? For example, how might a daily update differ from a weekly summary?
\end{enumerate}
\end{framed}

\section{Verbal Comprehension and Provenance Assessment}
\label{appendix:comprehension}

The verbal comprehension assessment was administered at the end of the session. Participants were asked to explain relevant system behaviors in their own words and, where applicable, to locate the corresponding information or controls in the interface. The prompts covered data provenance, AI verification evidence, and privacy and access controls. The prompts below were not administered verbatim; they were paraphrased for participants.

\begin{framed}
\subsection{Part A: Provenance and Data Origin}

\begin{enumerate}
    \item \textbf{Access-log identification:} Looking at your Access Logs screen, can you identify the most recent time [Care Actor Name] accessed your health metrics? Which information does the entry indicate was accessed?
    \item \textbf{Data-source identification:} In the 24-hour summary, point to the indicator showing whether the displayed heart-rate measurement came from your Samsung Galaxy Watch Ultra or from a manual entry. How can you tell?
\end{enumerate}

\subsection{Part B: AI Verification and Evidence Attribution}

\begin{enumerate}
    \item \textbf{Citation identification:} Point to a source citation used by the AI Verification Engine for the health tip. If you opened that citation, what information would you expect to find?
    \item \textbf{Confidence interpretation:} Suppose the Verification Engine marks a health tip as having low confidence but provides two supporting citations. How would you interpret this result? What would you examine before deciding whether to rely on the health tip?
\end{enumerate}

\subsection{Part C: Time-Bounded Access and Revocation}

\begin{enumerate}
    \item \textbf{Time-bounded access:} Suppose you granted [Care Actor Name] access to your health information for 24 hours. What would you expect to happen to their dashboard after the 24-hour period expires?
    \item \textbf{Immediate revocation:} As the Subject, show where you would go to revoke a Care Actor's access immediately. What would happen to the Care Actor's access after you revoke it?
\end{enumerate}
\end{framed}

The assessment was used to elicit participants' explanations and demonstrate how they navigated the relevant interface elements. Responses were considered alongside the walkthrough observations and interview accounts; they were not converted into a separate numerical comprehension score.

\end{document}